\documentclass[]{aastex62}
\usepackage{graphicx}
\usepackage{bmpsize}
\usepackage[space]{grffile}
\usepackage{latexsym}
\usepackage[caption=false]{subfig}
\usepackage{amsfonts,amsmath,amssymb}
\usepackage{url}
\usepackage[utf8]{inputenc}
\usepackage{fancyref}
\usepackage{lipsum}
\usepackage{ulem} 
\usepackage{color} 
\usepackage{comment}
\usepackage{dcolumn}
\usepackage{multirow}
\usepackage{bm}

\shorttitle{OB130911}
\shortauthors{Miyazaki et al.}
\accepted{December, 19, 2019}
\submitjournal{AJ}

\begin{document}
\title{OGLE-2013-BLG-0911Lb: A Secondary on the Brown-Dwarf Planet Boundary around an M-dwarf}

\correspondingauthor{Shota Miyazaki}
\email{miyazaki@iral.ess.sci.osaka-u.ac.jp}

\author[0000-0001-9818-1513]{Shota Miyazaki$^{\dag}$}
\affil{MOA collaboration}
\affil{Department of Earth and Space Science, Graduate School of Science, Osaka University, 1-1 Machikaneyama, Toyonaka, Osaka 560-0043, Japan}
\author{Takahiro Sumi}
\affil{MOA collaboration}
\affil{Department of Earth and Space Science, Graduate School of Science, Osaka University, 1-1 Machikaneyama, Toyonaka, Osaka 560-0043, Japan}
\author{David P. Bennett}
\affil{MOA collaboration}
\affil{Department of Physics, University of Notre Dame, Notre Dame, IN 46556, USA} 
\affil{Laboratory for Exoplanets and Stellar Astrophysics, NASA/Goddard Space Flight Center, Greenbelt, MD 20771, USA}
\author{Andrzej Udalski}
\affil{Warsaw University Observatory, Al.~Ujazdowskie~4, 00-478~Warszawa, Poland}
\affil{OGLE collaboration}
\author{Yossi Shvartzvald}
\affil{IPAC, Mail Code 100-22, Caltech, 1200 East California Boulevard, Pasadena, CA 91125, USA}
\affil{Wise team}
\author{Rachel Street} 
\affil{RoboNet}
\affil{Las Cumbres Observatory Global Telescope Network, 6740 Cortona Drive, suite 102, Goleta, CA 93117, USA}
\author{Valerio Bozza}
\affil{MiNDSTEp}
\affil{Dipartimento di Fisica "E.R. Caianiello,"Universit\'{a} di Salerno, Via Giovanni Paolo II 132, I-84084, Fisciano, Italy}
\affil{Istituto Nazionale di Fisica Nucleare, Sezione di Napoli, Napoli, Italy}
\author{Jennifer C. Yee}
\affil{Center for Astrophysics $|$ Harvard \& Smithsonian, 60 Garden St., Cambridge, MA 02138, USA}
\affil{$\mu$FUN}
\author{Ian A. Bond}
\affil{MOA collaboration}
\affil{Institute of Information and Mathematical Sciences, Massey University, Private Bag 102-904, North Shore Mail Centre, Auckland, New Zealand} 
\author{Nicholas Rattenbury}
\affil{MOA collaboration}
\affil{Department of Physics, University of Auckland, Private Bag 92019, Auckland, New Zealand}
\author{Naoki Koshimoto}
\affil{MOA collaboration}
\affil{Department of Earth and Space Science, Graduate School of Science, Osaka University, 1-1 Machikaneyama, Toyonaka, Osaka 560-0043, Japan}
\author{Daisuke Suzuki}
\affil{MOA collaboration}
\affil{Institute of Space and Astronautical Science, Japan Aerospace Exploration Agency, 3-1-1 Yoshinodai, Chuo, Sagamihara, Kanagawa 252-5210, Japan}
\author{Akihiko Fukui}
\affil{MOA collaboration}
\affil{Department of Earth and Planetary Science, Graduate School of Science, The University of Tokyo, 7-3-1 Hongo, Bunkyo-ku, Tokyo 113-0033, Japan}
\affil{Instituto de Astrof\'isica de Canarias, V\'ia L\'actea s/n, E-38205 La Laguna, Tenerife, Spain}
\nocollaboration

\author{F. Abe}
\affil{Institute for Space-Earth Environmental Research, Nagoya University, Nagoya 464-8601, Japan} 
\author{A. Bhattacharya}
\affil{Department of Physics, University of Notre Dame, Notre Dame, IN 46556, USA} 
\affil{Laboratory for Exoplanets and Stellar Astrophysics, NASA/Goddard Space Flight Center, Greenbelt, MD 20771, USA} 
\author{R. Barry}
\affil{Laboratory for Exoplanets and Stellar Astrophysics, NASA/Goddard Space Flight Center, Greenbelt, MD 20771, USA}
\author{M. Donachie}
\affil{Department of Physics, University of Auckland, Private Bag 92019, Auckland, New Zealand}
\author{H. Fujii}
\affil{Institute for Space-Earth Environmental Research, Nagoya University, Nagoya 464-8601, Japan} 
\author{Y. Hirao}
\affil{Department of Earth and Space Science, Graduate School of Science, Osaka University, 1-1 Machikaneyama, Toyonaka, Osaka 560-0043, Japan}
\author{Y. Itow}
\affil{Institute for Space-Earth Environmental Research, Nagoya University, Nagoya, 464-8601, Japan}
\author{Y. Kamei}
\affil{Institute for Space-Earth Environmental Research, Nagoya University, Nagoya 464-8601, Japan} 
\author{I. Kondo}
\affil{Department of Earth and Space Science, Graduate School of Science, Osaka University, 1-1 Machikaneyama, Toyonaka, Osaka 560-0043, Japan}
\author{M. C. A. Li}
\affil{Department of Physics, University of Auckland, Private Bag 92019, Auckland, New Zealand}
\author{C. H. Ling}
\affil{Institute of Information and Mathematical Sciences, Massey University, Private Bag 102-904, North Shore Mail Centre, Auckland, New Zealand}
\author{Y. Matsubara}
\affil{Institute for Space-Earth Environmental Research, Nagoya University, Nagoya, 464-8601, Japan}
\author{T. Matsuo}
\affil{Department of Earth and Space Science, Graduate School of Science, Osaka University, 1-1 Machikaneyama, Toyonaka, Osaka 560-0043, Japan}
\author{Y. Muraki}
\affil{Institute for Space-Earth Environmental Research, Nagoya University, Nagoya, 464-8601, Japan}
\author{M. Nagakane}
\affil{Department of Earth and Space Science, Graduate School of Science, Osaka University, 1-1 Machikaneyama, Toyonaka, Osaka 560-0043, Japan}
\author{K. Ohnishi}
\affil{Nagano National College of Technology, Nagano 381-8550, Japan}
\author{C. Ranc}
\affil{Laboratory for Exoplanets and Stellar Astrophysics, NASA/Goddard Space Flight Center, Greenbelt, MD 20771, USA}
\author{T. Saito}
\affil{Tokyo Metropolitan College of Industrial Technology, Tokyo 116-8523, Japan}
\author{A. Sharan}
\affil{Department of Physics, University of Auckland, Private Bag 92019, Auckland, New Zealand}
\author{H. Shibai}
\affil{Department of Earth and Space Science, Graduate School of Science, Osaka University, 1-1 Machikaneyama, Toyonaka, Osaka 560-0043, Japan}
\author{H. Suematsu}
\affil{Department of Earth and Space Science, Graduate School of Science, Osaka University, 1-1 Machikaneyama, Toyonaka, Osaka 560-0043, Japan} 
\author{D.J. Sullivan}
\affil{School of Chemical and Physical Sciences, Victoria University, Wellington, New Zealand}
\author{P. J. Tristram}
\affil{University of Canterbury Mt. John Observatory, P.O. Box 56, Lake Tekapo 8770, New Zealand}
\author{T. Yamakawa}
\affil{Institute for Space-Earth Environmental Research, Nagoya University, Nagoya 464-8601, Japan} 
\author{A. Yonehara}
\affil{Department of Physics, Faculty of Science, Kyoto Sangyo University, Kyoto 603-8555, Japan}
\collaboration{(MOA collaboration)}

\author{J. Skowron}
\affil{Warsaw University Observatory, Al.~Ujazdowskie~4, 00-478~Warszawa, Poland}
\author{R. Poleski}
\affil{Department of Astronomy, Ohio State University, 140 W. 18th Ave., Columbus, OH  43210, USA}
\author{P. Mr\'oz}
\affil{Warsaw University Observatory, Al.~Ujazdowskie~4, 00-478~Warszawa, Poland}
\affil{Division of Physics, Mathematics, and Astronomy, California Institute of Technology, Pasadena, CA 91125, USA}
\author{M. K. Szyma\'nski}
\affil{Warsaw University Observatory, Al.~Ujazdowskie~4, 00-478~Warszawa, Poland}
\author{I. Soszy{\'n}ski}
\affil{Warsaw University Observatory, Al.~Ujazdowskie~4, 00-478~Warszawa, Poland}
\author{P. Pietrukowicz}
\affil{Warsaw University Observatory, Al.~Ujazdowskie~4, 00-478~Warszawa, Poland}
\author{S. Koz\L owski}
\affil{Warsaw University Observatory, Al.~Ujazdowskie~4, 00-478~Warszawa, Poland}
\author{K. Ulaczyk}
\affil{Warsaw University Observatory, Al.~Ujazdowskie~4, 00-478~Warszawa, Poland}
\author{{\L}. Wyrzykowski}
\affil{Warsaw University Observatory, Al.~Ujazdowskie~4, 00-478~Warszawa, Poland}
\collaboration{(OGLE collaboration)}

\author{Matan Friedmann}
\affil{School of Physics and Astronomy and Wise Observatory, Tel-Aviv University, Tel-Aviv 6997801, Israel}
\author{Shai Kaspi}
\affil{School of Physics and Astronomy and Wise Observatory, Tel-Aviv University, Tel-Aviv 6997801, Israel}
\author{Dan Maoz}
\affil{School of Physics and Astronomy and Wise Observatory, Tel-Aviv University, Tel-Aviv 6997801, Israel}
\collaboration{(Wise team)}

\author{M. Albrow} 
\affil{Department of Physics and Astronomy, University of Canterbury, Private Bag 4800, Christchurch, New Zealand}
\author{G. Christie} 
\affil{Auckland Observatory, Auckland, New Zealand}
\author{D. L. DePoy} 
\affil{Department of Physics and Astronomy, Texas A\&M University, College Station, TX 77843-4242, USA}
\author{A. Gal-Yam} 
\affil{Department of Particle Physics and Astrophysics, Weizmann Institute of Science, 76100 Rehovot, Israel}
\author{A. Gould}
\affil{Department of Astronomy, Ohio State University, 140 West 18th Avenue, Columbus, OH 43210, USA}
\affil{Korea Astronomy and Space Science Institute, Daejon 34055, Republic of Korea}
\affil{Max Planck Institute for Astronomy, K\"{o}nigstuhl 17, D-69117 Heidelberg, Germany}
\author{C.-U. Lee} 
\affil{Korea Astronomy and Space Science Institute, 776 Daedukdae-ro, Daejeon, Korea}
\affil{University of Science and Technology, Korea, (UST), 217 Gajeong-ro Yuseong-gu, Daejeon 34113, Korea}
\author{I. Manulis} 
\affil{Department of Particle Physics and Astrophysics, Weizmann Institute of Science, 76100 Rehovot, Israel}
\author{J. McCormick} 
\affil{Farm Cove Observatory, Centre for Backyard Astrophysics, Pakuranga, Auckland, New Zealand}
\author{T. Natusch} 
\affil{ Auckland Observatory, Auckland, New Zealand}
\affil{Institute for Radio Astronomy and Space Research (IRASR), AUT University, Auckland, New Zealand}
\author{H. Ngan} 
\affil{Auckland Observatory, Auckland, New Zealand}
\author{R. W. Pogge}
\affil{Department of Astronomy, The Ohio State University, 140 W 18th Ave., Columbus, OH, 43210}
\affil{Center for Cosmology \& AstroParticle Physics, The Ohio State University, 191 West Woodruff Avenue, Columbus, OH 43210}
\author{I. Porritt}
\affil{Turitea Observatory, Palmerston North, New Zealand}
\collaboration{($\mu$FUN)}

\author{Y. Tsapras}
\affil{Astronomisches Rechen-Institut, Zentrum f\"{u}r Astronomie der Universit\"{a}t Heidelberg (ZAH), D-69120 Heidelberg, Germany}
\author{E. Bachelet}
\affil{Las Cumbres Observatory Global Telescope Network, 6740 Cortona Drive, suite 102, Goleta, CA 93117, USA}
\affil{Qatar Environment and Energy Research Institute(QEERI), HBKU, Qatar Foundation, Doha, Qatar}
\author{M.P.G. Hundertmark}
\affil{Astronomisches Rechen-Institut, Zentrum f\"{u}r Astronomie der Universit\"{a}t Heidelberg (ZAH), D-69120 Heidelberg, Germany}
\author{M. Dominik}
\affil{Centre for Exoplanet Science, SUPA School of Physics \& Astronomy, University of St Andrews, North Haugh, St Andrews, KY16 9SS, United Kingdom}
\author{D. M. Bramich}
\affil{Center for Space Science, NYUAD Institute, New York University Abu Dhabi, PO Box 129188, Saadiyat Island, Abu Dhabi, UAE}
\affil{Center for Astro, Particle and Planetary Physics, New York University Abu Dhabi, PO Box 129188, Saadiyat Island, Abu Dhabi, UAE}
\affil{Division of Engineering, New York University Abu Dhabi, PO Box 129188, Saadiyat Island, Abu Dhabi, UAE}
\author{A. Cassan}
\affil{Institut d’Astrophysique de Paris, Sorbonne Universit\'e, CNRS, UMR 7095, 98 bis bd Arago, 75014 Paris, France}
\author{R. Figuera Jaimes}
\affil{Centre for Exoplanet Science, SUPA School of Physics \& Astronomy, University of St Andrews, North Haugh, St Andrews, KY16 9SS, United Kingdom}
\affil{European Southern Observatory, Karl-Schwarzschild-Str. 2, D-85748 Garching bei M\"{u}nchen, Germany}
\author{K. Horne}
\affil{Centre for Exoplanet Science, SUPA School of Physics \& Astronomy, University of St Andrews, North Haugh, St Andrews, KY16 9SS, United Kingdom}
\author{R. Schmidt}
\affil{Astronomisches Rechen-Institut, Zentrum f\"{u}r Astronomie der Universit\"{a}t Heidelberg (ZAH), D-69120 Heidelberg, Germany}
\author{C. Snodgrass}
\affil{Institute for Astronomy, University of Edinburgh, Royal Observatory, Blackford Hill, Edinburgh, EH9 3HJ, U.K.}
\author{J. Wambsganss}
\affil{Astronomisches Rechen-Institut, Zentrum f\"{u}r Astronomie der Universit\"{a}t Heidelberg (ZAH), D-69120 Heidelberg, Germany}
\author{I. A. Steele}
\affil{Astrophysics Research Institute, Liverpool John Moores University, Liverpool CH41 1LD, UK}
\author{J. Menzies}
\affil{South African Astronomical Observatory, P.O. Box 9, Observatory 7935, South Africa}
\author{S. Mao}
\affil{Physics Department and Tsinghua Centre for Astrophysics, Tsinghua University, Beijing 100084, China}
\affil{National Astronomical Observatories, Chinese Academy of Sciences, 20A Datun Road, Chaoyang District, Beijing 100012, China}
\affil{ Jodrell Bank Centre for Astrophysics, School of Physics and Astronomy, The University of Manchester, Oxford Road, Manchester M13 9PL, UK}
\collaboration{(RoboNet)}

\author{U. G. J\o rgensen}
\affil{Niels Bohr Institute \& Centre for Star and Planet Formation, University of Copenhagen, \O ster Voldgade 5, DK-1350 Copenhagen, Denmark}
\author{M. J. Burgdorf}
\affil{Meteorologisches Institut, Universit\"{a}t Hamburg, Bundesstra\ss e 55, D-20146 Hamburg, Germany}
\author{S. Ciceri}
\affil{Max Planck Institute for Astronomy, K\"{o}nigstuhl 17, D-69117 Heidelberg, Germany}
\author{S. Calchi Novati}
\affil{IPAC, Mail Code 100-22, Caltech, 1200 E. California Blvd., Pasadena, CA 91125, USA}
\author{G. D'Ago}
\affil{Dipartimento di Fisica "E.R. Caianiello,"Universit\'{a} di Salerno, Via Giovanni Paolo II 132, I-84084, Fisciano, Italy}
\affil{Istituto Nazionale di Fisica Nucleare, Sezione di Napoli, Napoli, Italy}
\affil{Spitzer Science Center, MS 220-6, California Institute of Technology, Pasadena, CA, USA}
\author{D. F. Evans}
\affil{Astrophysics Group, Keele University, Staffordshire, ST5 5BG, UK}
\author{T. C. Hinse}
\affil{Korea Astronomy \& Space Science Institute, 776 Daedukdae-ro, Yuseong-gu, 305-348 Daejeon, Korea}
\author{N. Kains}
\affil{Jodrell Bank Centre for Astrophysics, School of Physics and Astronomy, The University of Manchester, Oxford Road, Manchester M13 9PL, UK}
\author{E. Kerins}
\affil{Jodrell Bank Centre for Astrophysics, School of Physics and Astronomy, The University of Manchester, Oxford Road, Manchester M13 9PL, UK}
\author{H. Korhonen}
\affil{Dark Cosmology Centre, Niels Bohr Institute, University of Copenhagen, Juliane Maries Vej 30, 2100 Copenhagen, Denmark}
\author{L. Mancini}
\affil{Max Planck Institute for Astronomy, K\"{o}nigstuhl 17, D-69117 Heidelberg, Germany}
\author{A. Popovas}
\affil{Niels Bohr Institute \& Centre for Star and Planet Formation, University of Copenhagen, \O ster Voldgade 5, DK-1350 Copenhagen, Denmark}
\author{M. Rabus}
\affil{Instituto de Astrof\'{\i}sica, Facultad de F\'{\i}sica, Pontificia Universidad Cat\'{o}lica de Chile, Av. Vicu\~{n}a Mackenna 4860, 7820436 Macul, Santiago, Chile}
\author{S. Rahvar}
\affil{Department of Physics, Sharif University of Technology, P.O. Box 11155-9161 Tehran, Iran}
\author{G. Scarpetta}
\affil{Dipartimento di Fisica ``E.R. Caianiello'', Universit\`a di Salerno, Via Giovanni Paolo II 132, Fisciano 84084, Italy.}
\affil{International Institute for Advanced Scientific Studies (IIASS), Via G. Pellegrino 19, 84019 Vietri sul Mare (SA), Italy.}
\author{J. Skottfelt}
\affil{Niels Bohr Institute \& Centre for Star and Planet Formation, University of Copenhagen, \O ster Voldgade 5, DK-1350 Copenhagen, Denmark}
\affil{Centre for Electronic Imaging, Department of Physical Sciences, The Open University, Milton Keynes, MK7 6AA, UK}
\author{J. Southworth}
\affil{Astrophysics Group, Keele University, Staffordshire, ST5 5BG, UK}
\author{G. D'Ago}
\affil{Instituto de Astrofisica, Facultad de Fisica, Pontificia Universidad Catolica de Chile, Av. Vicuna Mackenna 4860,7820436 Macul, Santiago, Chile}
\author{N. Peixinho}
\affil{CITEUC—Centre for Earth and Space Science Research of the University of Coimbra, Observat\'{o}rio Geof\'{\i}sico e Astron\'{o}mico da U.C., 3030-004 Coimbra, Portugal}
\affil{Unidad de Astronom\'{\i}a, Fac. de Cs. B\'{a}sicas, Universidad de Antofagasta, Avda U. de Antofagasta 02800, Antofagasta, Chile}
\author{P. Verma}
\affil{Istituto Internazionale per gli Alti Studi Scientifici (IIASS), Via G. Pellegrino 19, I-84019 Vietri sul Mare (SA), Italy}
\collaboration{(MiNDSTEp)}

\begin{abstract}
We present the analysis of the binary-lens microlensing event OGLE-2013-BLG-0911.
The best-fit solutions indicate the binary mass ratio of $q\simeq0.03$ which differs from that reported in \citet{Shvartzvald+2016}.
The event suffers from the well-known close/wide degeneracy, resulting in two groups of solutions for the projected separation normalized by the Einstein radius of $s\sim0.15$ or $s\sim7$.
The finite source and the parallax observations allow us to measure the lens physical parameters. 
The lens system is an M-dwarf orbited by a massive Jupiter companion at very close ($M_{\rm host}=0.30^{+0.08}_{-0.06}M_{\odot}$, $M_{\rm comp}=10.1^{+2.9}_{-2.2}M_{\rm Jup}$, $a_{\rm exp}=0.40^{+0.05}_{-0.04}{\rm au}$) or wide ($M_{\rm host}=0.28^{+0.10}_{-0.08}M_{\odot}$, $M_{\rm comp}=9.9^{+3.8}_{-3.5}M_{\rm Jup}$, $a_{\rm exp}=18.0^{+3.2}_{-3.2}{\rm au}$) separation.
Although the mass ratio is slightly above the planet-brown dwarf (BD) mass-ratio boundary of $q=0.03$ which is generally used, the median physical mass of the companion is slightly below the planet-BD mass boundary of $13M_{\rm Jup}$. 
It is likely that the formation mechanisms for BDs and planets are different and the objects near the boundaries could have been formed by either mechanism. 
It is important to probe the distribution of such companions with masses of $\sim13M_{\rm Jup}$ in order to statistically constrain the formation theories for both BDs and massive planets. 
In particular, the microlensing method is able to probe the distribution around low-mass M-dwarfs and even BDs which is challenging for other exoplanet detection methods.
\end{abstract} 
\keywords{microlensing --- exoplanets --- brown dwarfs}

\section{Introduction}\label{sec:Introduction}
\begin{figure}[htb!]
\centering
\includegraphics[scale=0.6]{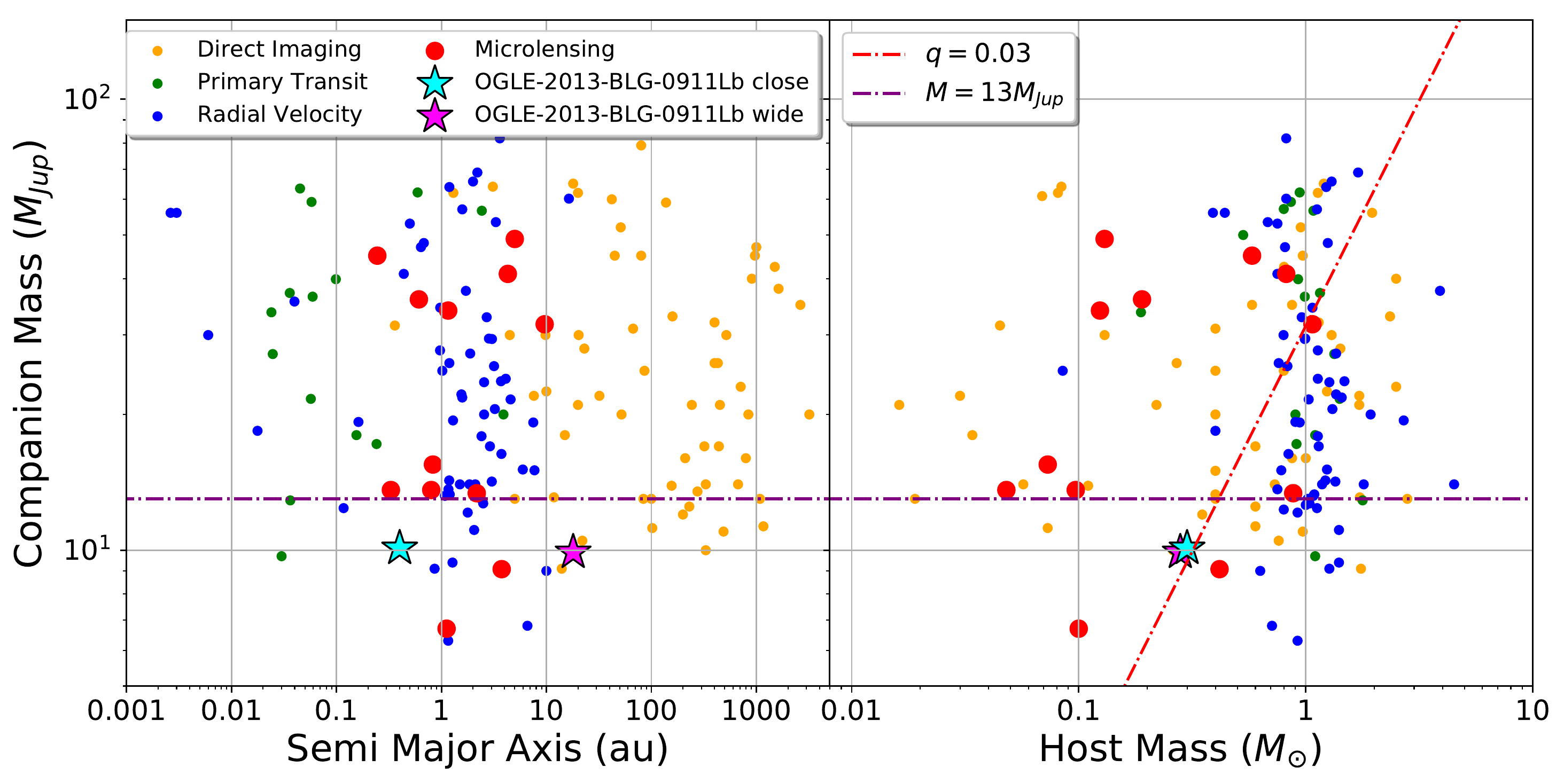}
\caption{
Distributions of discovered BD/massive-planet companions ($5M_{\rm Jup}\le M\le75M_{\rm Jup}$) obtained from http://exoplanet.eu, in which the vertical axis shows the companion masses.
The horizontal axes for the left and right panels indicate the semi major axes and host masses, respectively.
The yellow, green, blue and red points indicate the BD/massive-planet companions discovered by Imaging, Transit, Radial Velocity and Microlensing method, respectively.
The two solutions for OGLE-2013-BLG-0911Lb are represented as stars.  
}
\label{fig:BD_dist}
\end{figure}

Brown dwarfs (BDs) have masses of $13-75M_{\rm Jup}$ being intermediate between the masses of the main-sequence stars and planets \citep{Burrows+1993}.
Although the existence of BDs was firstly proposed in \citet{Kumar1962}, there had been no observational evidence for BDs until 1995 \citep{Nakajima+1995} owing to their low luminosities and temperatures.
To date, more than ten thousand field BDs have been discovered by several survey groups, which are summarized in the Table 1 of \citet{Carnero Rosell+2019}.
Most current theories predict that field BDs are formed in a fashion similar to that of main sequence stars, through direct gravitational collapse and turbulent fragmentation of molecular clouds \citep{Luhman2012}. 
These theories are observationally supported. 
For example, \citet{Andre+2012} found self-gravitating dense clumps of gasses and dust with mass  0.015-0.03$M_{\odot}$ ,which are similar to those of low mass BDs.
On the other hand,  the core accretion mechanism \citep{Mordasini+2009,Tanigawa+2016} and that of gravitational instability \citep{Boss1997,Boss2001} are also able to produce companions of BD masses in protoplanetary disks.
Radial velocity (RV) surveys have revealed that the frequency of BD companions with orbital radii less than $\sim3$ au around main sequence stars is relatively lower than that of stellar and planetary-mass companions \citep{Marcy+2000, Grether+2006,Johnson+2010}, the so-called ``brown dwarf desert''.  
It is likely that this BD deficit is because of differences between the formation mechanisms of companions with  planetary mass and stellar mass. 
However, it is not yet clear if the BD-mass companions formed like planets in the protoplanetary disk,  formed as binary stars in the molecular cloud or were captured by the primary stars.
Some theories have suggested that the BD desert might be an outcome of the interaction between massive companions and protoplanetary disks and/or of tidal evolution \citep{Armitage+2002, Matzner+2005,Duchene+2013}.
\vspace{0.1in}

Gravitational microlensing \citep{Mao+1991} surveys have probed the distribution of the outer planetary systems beyond the snow line \citep{Hayashi1981}, where the ice-dominated solid materials are rich, leading to efficient formation of gas-giant planets according to the core accretion theory \citep{Lissauer1993,Pollack+1996}.
Because microlensing does not depend on the luminosity of the host star, the technique is sensitive to companions to low mass objects such as late M-dwarfs or even BDs. 
Furthermore, the host and any companions can still be inferred at distances all the way to the Galactic bulge.
In contrast, the RV and transit \citep{Borucki+2010} methods, which have discovered the bulk of currently known exoplanets and BDs orbiting around hosts, have only a sensitivity to companions relatively close to hosts and whose hosts are sufficiently bright.
Figure \ref{fig:BD_dist} shows the distribution of discovered BD/massive-planet companions around main sequence stars and BDs. 
The RV (blue dots) and transit (green dots) methods have discovered a lot of the companions around 1 $M_{\odot}$ stars but only a few around low-mass stars below 0.5 $M_{\odot}$.
This would be caused by an observational bias due to the faintness of low-mass stars in visible wavelength range. 
The direct imaging (orange dots) method has detected the companions around hosts with masses of $0.01-3\;M_{\odot}$ but it could not have resolved the companions with relatively short orbital radii.
On the other hand, microlensing (red dots) has discovered BD/massive-planet companions around hosts with masses of $\sim0.05-1M_{\odot}$ with orbital radii of $\sim0.3-10$ au (e.g. \citealp{Ranc+2015,Han+2017,Ryu+2018}), which are complementary to other detection methods. 
\citet{Gaudi2002} estimated that more than 25\% of BD companions with separations $\sim1-10$ au would be detected by present microlensing surveys. 
According to the standard core accretion theory, massive planets and also BDs are more difficult to form around low-mass M dwarfs than solar-type stars owing to low disk surface densities \citep{Ida+2005} and long timescales \citep{Laughlin+2004}.
It is possible to constrain the BD formation mechanism around late M dwarfs from a statistical analysis of microlensing results in the BD-mass regime, which can be compared to the lack of close-in BD companions around solar-type stars found by RV observations.
\vspace{0.1in}

\citet{Shvartzvald+2016}, hereafter S16, conducted a statistical analysis of the first four seasons of a ``second-generation'' microlensing survey \citep{Gaudi+2009} which consisted of the observations by the Optical Gravitational Lensing Experiment (OGLE; \citealp{Udalski+1994}) collaboration, the Microlensing Observations in Astrophysics (MOA; \citealp{Bond+2001,Sumi+2003}) collaboration and the Wise team \citep{Shvartzvald+2012}.
They analyzed 224 microlensing events and found 29 ``anomalous'' events which imply the presence of a companion to the lens host. 
They performed an automated coarse grid search for light curve modelings rather than a detailed modeling of individual events for their statistical study.
Finally, they derived the planet (binary) frequency distribution as a function of companion-to-host mass ratio $q$ and found a possible deficit at $q\sim10^{-2}$.
However, it is worthwhile to conduct the detailed analysis of  individual ``planetary candidate'' in their sample which do not have any models in literature. 
For example, they reported that OGLE-2013-BLG-0911 has a planetary mass-ratio of $q\approx3\times10^{-4}$,  but we found new preferred solutions with a less extreme mass ratio, $q\approx3\times10^{-2}$.
\vspace{0.1in}

Here, we present the analysis of a high-magnification (maximum magnification of $A_{\rm max}\sim220$) microlensing event, OGLE-2013-BLG-0911. 
The ``anomaly'' due to a companion to the lens star was clearly detected near the peak of the light curve.
We present the observations and datasets of the event in Section \ref{sec:Observation}.
Our light curve analysis is described in Section \ref{sec:Modeling}.
In Section \ref{sec:source}, we present our analysis of the source properties.
The physical parameters of the lens system are described in Section \ref{sec:lens_priority}. 
We summarize and discuss the result in Section \ref{sec:sam_dis}.

\section{Observation \& Data Sets \label{sec:Observation}}
\begin{deluxetable*}{lllccllll}
\tablecaption{Data Sets for OGLE-2013-BLG-0911}
\tablecolumns{9}
\tabletypesize{\footnotesize}
\tablehead{
\multicolumn{1}{l}{Site} & \multicolumn{1}{l}{Telescope} & \multicolumn{1}{l}{Collaboration} & \colhead{Label} & \colhead{Filter} &
\colhead{$N_{\rm use}$} & \colhead{$k$\tablenotemark{$a$}} 
}
\startdata
Mount John Observatory & MOA-II 1.8m & MOA & MOA & $Red$ & 8761 & 1.055 \\
Las Campanas Observatory & Warsaw 1.3m & OGLE & OGLE & $I$ & 6895 & 1.480 \\
Las Campanas Observatory & Warsaw 1.3m & OGLE & OGLE & $V$ & 78 & 1.344 \\
Florence and George Wise Observatory & Wise 1m & Wise & Wise1m & $I$ & 253 & 0.947  \\
Cerro Tololo-Inter American Observatory (CTIO) & SMARTS 1.3m & $\mu$FUN & CT13 & $I$ & 189 & 1.230 \\
Cerro Tololo-Inter American Observatory (CTIO) & SMARTS 1.3m & $\mu$FUN & CT13 & $V$ & 35 & 1.182 \\
Farm Cove Observatory & Farm Cove 0.36m & $\mu$FUN & FCO & Unfiltered & 55 & 2.146 \\
Weizmann Institute of Science, Marty S. Kraar Observatory & Weizmann 16inch & $\mu$FUN & WIS & $I$ & 17 & 1.140\\
Haleakala Observatory & Faulkes North 2.0m & RoboNet & FTN & $i^\prime$ & 27 & 2.181 \\
Siding Spring Observatory (SSO) & LCO 1.0m, Dome A & RoboNet & cojA & $i^\prime$ & 31 & 1.920 \\ 
Cerro Tololo Inter-American Observatory (CTIO) & LCO 1.0m, Dome B & RoboNet & lscB & $i^\prime$ & 51 & 1.311 \\ 
Cerro Tololo Inter-American Observatory (CTIO) & LCO 1.0m, Dome C & RoboNet & lscC & $i^\prime$ & 71 & 2.315 \\ 
South African Astronomical Observatory (SAAO) & LCO 1.0m, Dome A & RoboNet & cptA & $i^\prime$ & 32 & 0.559 \\
South African Astronomical Observatory (SAAO) & LCO 1.0m, Dome B & RoboNet & cptB & $i^\prime$ & 8 & 0.497 \\
ESO's La Silla Observatory & Danish 1.54m & MiNDSTEp & Dan & $I$ & 76 & 2.087 \\
Salerno University Observatory & Salerno 0.36m & MiNDSTEp & Sal & $I$ & 20 & 1.607 \\
\enddata
\label{table:data}
\tablecomments{The WIS, Sal and lscC data are binned for 0.01 days.}
\tablenotetext{a}{The coefficient for error renormalization, see text.}
\end{deluxetable*}

\subsection{Observation}
The microlensing event OGLE-2013-BLG-0911 was discovered and alerted as a microlensing candidate on 2013 June 3 UT 21:51 by the fourth phase of the OGLE collaboration (OGLE-IV; \citealp{Udalski+2015}).
OGLE-IV\footnote{http://ogle.astrouw.edu.pl/ogle4/ews/ews.html} is conducting a microlensing exoplanet search toward the Galactic bulge using the 1.3m Warsaw telescope of Las Campanas Observatory in Chile with a wide total field of view (FOV) of 1.4 deg$^{2}$.
The OGLE observations were conducted using the standard $I$- and near-standard $V$-band filters.
The second phase of the MOA collaboration\footnote{https://www.massey.ac.nz/~iabond/moa/alerts/} (MOA-II; \citealp{Bond+2017}) is also carrying out a microlensing survey toward the Galactic bulge using the 1.8m MOA-II telescope with a 2.2deg$^{2}$ FOV CCD camera (MOA-cam3; \citealp{Sako+2008}) at Mount John Observatory (MJO) in New Zealand.
Thanks to its wide FOV, the MOA collaboration is observing bulge stars with a cadence of 15-90min every day depending on the field.
The MOA survey independently discovered and issued an alert for the event as MOA-2013-BLG-551.
The MOA observations were conducted using a custom wide-band filter, ``MOA-Red'', which corresponds approximately to the combination of the standard $I$ and $R$ filters.
The Wise\footnote{http://wise-obs.tau.ac.il/~wingspan/} team also conducted a microlensing survey from 2010 to 2015 and monitored a field of 8 ${\rm deg}^2$ within the observational footprints of both OGLE and MOA \citep{Shvartzvald+2012}.
They observed using the 1m Wise telescope at Wise Observatory in Israel with a 1 ${\rm deg}^2$ FOV LAIWO camera \citep{Gorbikov+2010} and the cadence for each of the eight Wise fields was $\sim30$min.
\vspace{0.1in}

The event was located at $({\rm R.A.}, {\rm Dec.})_{\rm J2000}$=(17:55:31.98, $-$29:15:13.8) or Galactic coordinates $(l, b)$ = (0.84$^{\circ}$, $-$2.02$^{\circ}$).
Real-time analysis predicted the event would reach high peak magnification during which the sensitivity to low-mass companions is high \citep{Griest+1998, Rattenbury+2002}. 
Follow-up observations during the period of high magnification were encouraged to capture short planetary signals.
Consequently, in addition to the OGLE and MOA survey observations,  the light curve was densely observed by several follow-up groups: Microlensing Follow Up Network ($\mu$FUN; \citealp{Gould+2006}), Microlensing Network for the Detection of Small Terrestrial Exoplanets (MiNDSTEp; \citealp{Dominik+2010}) and RoboNet \citep{Tsapras+2009, Dominik+2019}.
Hereafter, we refer this event as OGLE-2013-BLG-0911.

\subsection{Data reduction}
All the datasets of OGLE-2013-BLG-0911 are summarized in Table \ref{table:data}.
Most photometric pipelines use the Difference Image Analysis (DIA; \citealp{Alard-Lupton1998, Alard2000}) technique, which is very effective in high stellar density fields such as those towards the Galactic bulge.
The MOA and $\mu$FUN CTIO data were reduced with the MOA implementation of the DIA method \citep{Bond+2001,Bond+2017}.
The OGLE data were reduced by OGLE's DIA pipeline \citep{Wozniak2000}.
The Wise data were reduced using the pySIS DIA software \citep{Albrow+2009}.
The other $\mu$FUN data and MiNDSTEp data were reduced by DoPhot \citep{Schechter+1993} and DanDIA \citep{Bramich2008,Bramich+2013}.
RoboNet data were reduced using a customized version of the DanDIA pipeline \citep{Bramich2008}.
\vspace{0.1in}

It is known that the nominal photometric error bars given by each photometric pipeline are potentially underestimated in high stellar density fields toward the bulge.
Therefore, we empirically renormalized the error bars for each data set following procedure of \citet{Bennett+2008} and \citet{Yee+2012}, i.e., 
\begin{equation} \label{eq:err}
\sigma^{'}_{i}=k\sqrt{\sigma^2_{i}+e^2_{\rm min}}\ ,
\end{equation}
where $\sigma^{'}_{i}$ and $\sigma_{i}$ represent the renormalized errors and the original errors given by the pipelines, respectively. 
The parameters $k$ and $e_{\rm min}$ are the coefficients for the error renormalization. 
Here, $e_{\rm min}$ represents the systematic errors when the source flux is significantly magnified. 
We added 0.3\% in quadrature to each error, i.e. $e_{\rm min}=0.003$, and then calculated $k$ values in order to achieve a value of $\chi^{2}/{\rm dof}=1$ for each dataset \citep{Bennett+2014,Skowron+2016}.
We list the renormalization coefficients $k$ in Table \ref{table:data} along with the number of used data points $N_{\rm use}$.
We confirmed that the final best-fit model is consistent with the preliminary best-fit model found using the datasets before the error renormalization. 
\vspace{0.1in}

\section{Light Curve Modeling \label{sec:Modeling}}

\begin{deluxetable}{lllcccccccc}
\tablecaption{Comparisons between each microlensing model\label{tab:chi2comp}}
\tablehead{
\multicolumn{3}{c}{Model}  & $N_{\rm param}$\tablenotemark{$a$} & $\chi^{2}$ &  BIC\tablenotemark{$b$} & $\Delta\chi^{2}$ & $\Delta$BIC
}
\startdata
1L1S  &  & Static & 4 & 21027.4 & 21066.3 & 4485.1 & 4368.5\\ \hline
1L2S  &  & Static & 10 & 18631.0 & 18728.2 & 2088.7 & 2030.4\\ 
1L2S  &  & Xallarap & 12 & 17554.3 & 17670.9 & 1212.0 & 973.1\\ \hline
2L1S  & $(s<1)$ & Static & 7 & 17473.7 & 17541.7 & 931.4 & 843.9\\
2L1S  & $(s<1, u_0>0)$ & Parallax & 9 & 17262.7 & 17350.2 & 720.4 & 652.4\\
2L1S  & $(s<1, u_0>0)$ & Xallarap & 14 & 16587.6 & 16723.6 & 45.3 & 25.8\\
2L1S  & $(s<1, u_0>0)$ & Parallax+Xallarap & 16 & 16558.9 & 16714.4 & 16.6 & 16.6\\ \hline
2L2S  & $(s<1, u_0>0)$ & Parallax+Xallarap & 16 & 16542.3 & 16697.8 & - & - \\ 
\enddata
\tablenotetext{a}{Number of fitting parameters.}
\tablenotetext{b}{Bayesian information criterion.}
\end{deluxetable}

\begin{figure}
\centering
\includegraphics[angle=-90, scale=0.5]{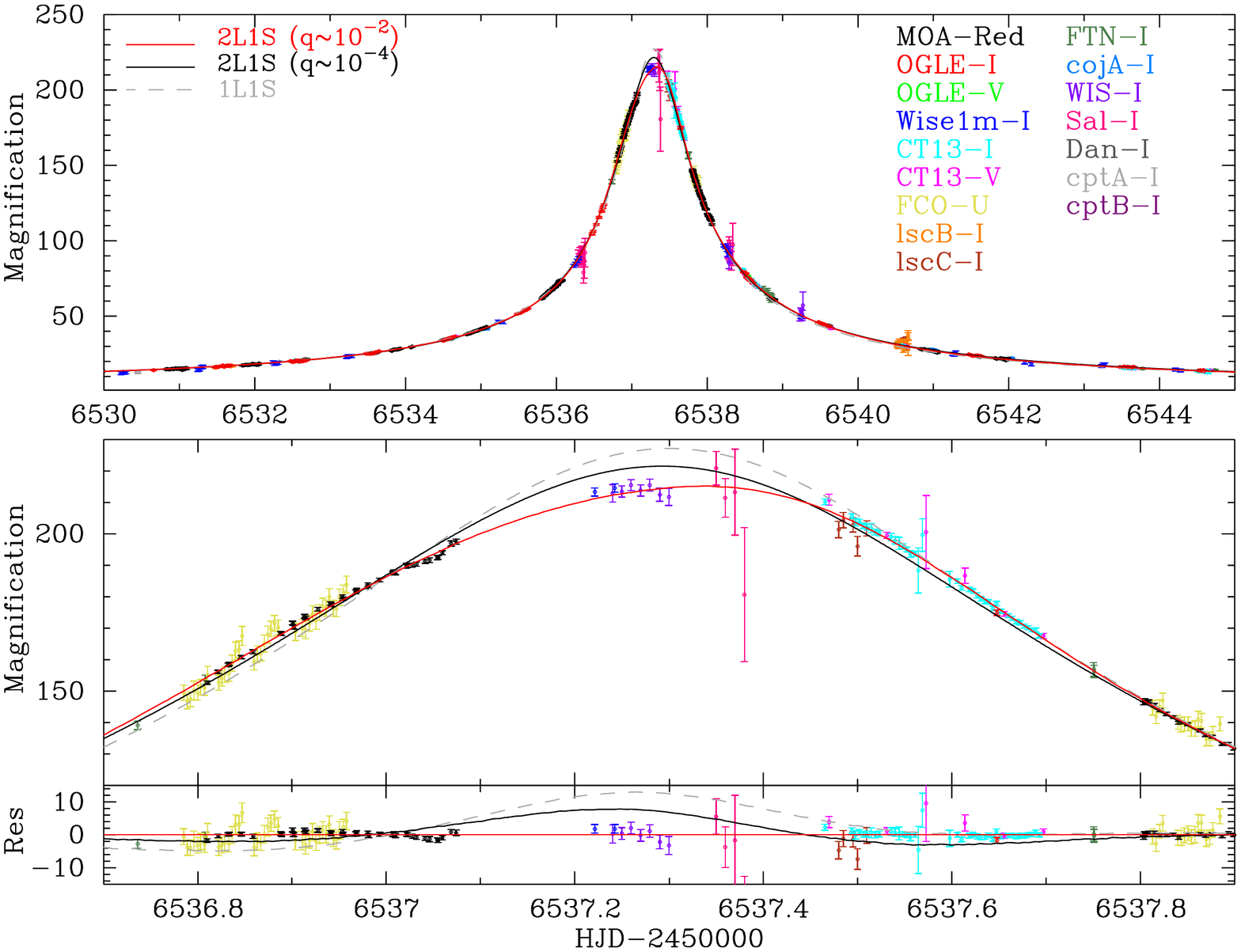}
\caption{
(Top) The light curve of OGLE-2013-BLG-0911.
Each color on the data point corresponds to each instrument, shown on the right.
The error bars are renormalized following Equation (\ref{eq:err}).
The solid red, black and dashed gray curves represent the static 2L1S with $q\sim10^{-2}$, 2L1S with $q\sim10^{-4}$ and 1L1S models, respectively.
(Middle) A zoom-in around the peak.
(Bottom) Residuals of the zoom-in light curve from the model of 2L1S with $q\sim10^{-2}$.
}
\label{fig:LC_MB13551}
\end{figure}

Here, we present the light curve modeling for OGLE-2013-BLG-0911.
Figure \ref{fig:LC_MB13551} represents the light curve of OGLE-2013-BLG-0911.
The main anomalous feature can be seen between $6536.8< {\rm HJD}-245000 < 6537.6$.
A standard single-lens single-source (1L1S) model fits the data worse than a binary-lens single-source (2L1S) model by $\Delta\chi^{2}>3500$.
In following sections, we present the details of the light curve modeling for OGLE-2013-BLG-0911.
In Table \ref{tab:chi2comp}, we summarize the comparisons of the $\chi^2$, number of fitting parameters and Bayesian information criterion (BIC) between microlensing models we examined. 
\vspace{0.1in}

\subsection{Model Description}
Assuming a single source star, the observed flux at any given time in a microlensing event, $F_{\rm obs}(t)$, can be modeled by the following equation,
\begin{equation}\label{eq:Flux}
F_{\rm obs}(t)= A(t)F_{\rm s}+F_{\rm b}\ ,
\end{equation}
where $A(t)$ is the magnification of the source flux, $F_{\rm s}$ is the unmagnified source flux, and $F_{\rm b}$ is the blend flux.
We note that $F_{\rm s}$ and $F_{\rm b}$ can be, during the fitting process, solved analytically by the linear equation (\ref{eq:Flux}) at given $A(t)$.
For a standard single-lens single-source (1L1S) model, there are four parameters that describe the light curve features \citep{Paczynski1986}; the time of the source approaching closest to the lens center of mass, $t_{0}$; the impact parameter, $u_{0}$, in unit of the angular Einstein radius, $\theta_{\rm E}$; the Einstein radius crossing time, $t_{\rm E}$; the source angular radius, $\rho$, in unit of $\theta_{\rm E}$.
The measurement of $\rho$ is important because it leads to a determination of $\theta_{\rm E}$ which is needed for the determination of the the mass-distance relation of the lens system. 
\vspace{0.1in}

In our fitting process, we used a Markov Chain Monte Carlo (MCMC) method \citep{Verde+2003} combined with our implementation of the inverse ray-shooting method \citep{Bennett+Rhie1996, Bennett2010} in order to find the best-fit model and estimate the parameter uncertainties from MCMC stationary distribution for each parameter.
Linear limb-darkening models were used to describe the source star(s) in this work.
From the measurement of the intrinsic source color of $(V-I)_{s,0}=0.71$ described in Section \ref{sec:source}, we assumed the effective temperature $T_{\rm eff}=5750$K \citep{Gonzalez+2009}, the surface gravity $\log g=4.5$ and metallicity log$[M/H]=0$.
According to the ATLAS model of \citet{Claret+2011}, we selected the limb darkening coefficients of $u_{Red}=0.5900$, $u_{I}=0.5493$, $u_{V}=0.7107$.
Here, $u_{Red}$ for the MOA-$Red$ band is estimated as the mean of  $u_{I}$ and $u_{R}$ and the $R$-band coefficient $u_{R}=0.6345$ is used for an unfiltered band.

\subsection{Binary Lens (2L1S) Model}
For a standard binary-lens single-source (2L1S) model, there are three additional parameters; the lens mass ratio between the host and a companion, $q$; the projected binary separation in unit of the Einstein radius, $s$; the angle between the source trajectory and the binary-lens axis, $\alpha$. 
Here, we introduce two fitting parameters $t_{c}$ and $u_{c}$, for wide ($s>1$) models.
If $s>1$, the system center in our numerical code is offset from the binary center of mass by 
\begin{equation}
\Delta(x,y)=\biggl[ \frac{q}{1+q}\left(\frac{1}{s}-s\right), 0\biggr] \nonumber
\end{equation}
where $(x,y)$ are the parallel and vertical coordinate axes to the binary-lens axis on the lens plane \citep{Skowron+2011}, and then we define the time of the source approaching closest to the ``system center'' and the impact parameter in units of the angular Einstein radius as $t_{c}$ and $u_{c}$, respectively.

\subsubsection{Static models}
\begin{figure}[htb!]
\centering
\includegraphics[scale=0.6,angle=-90]{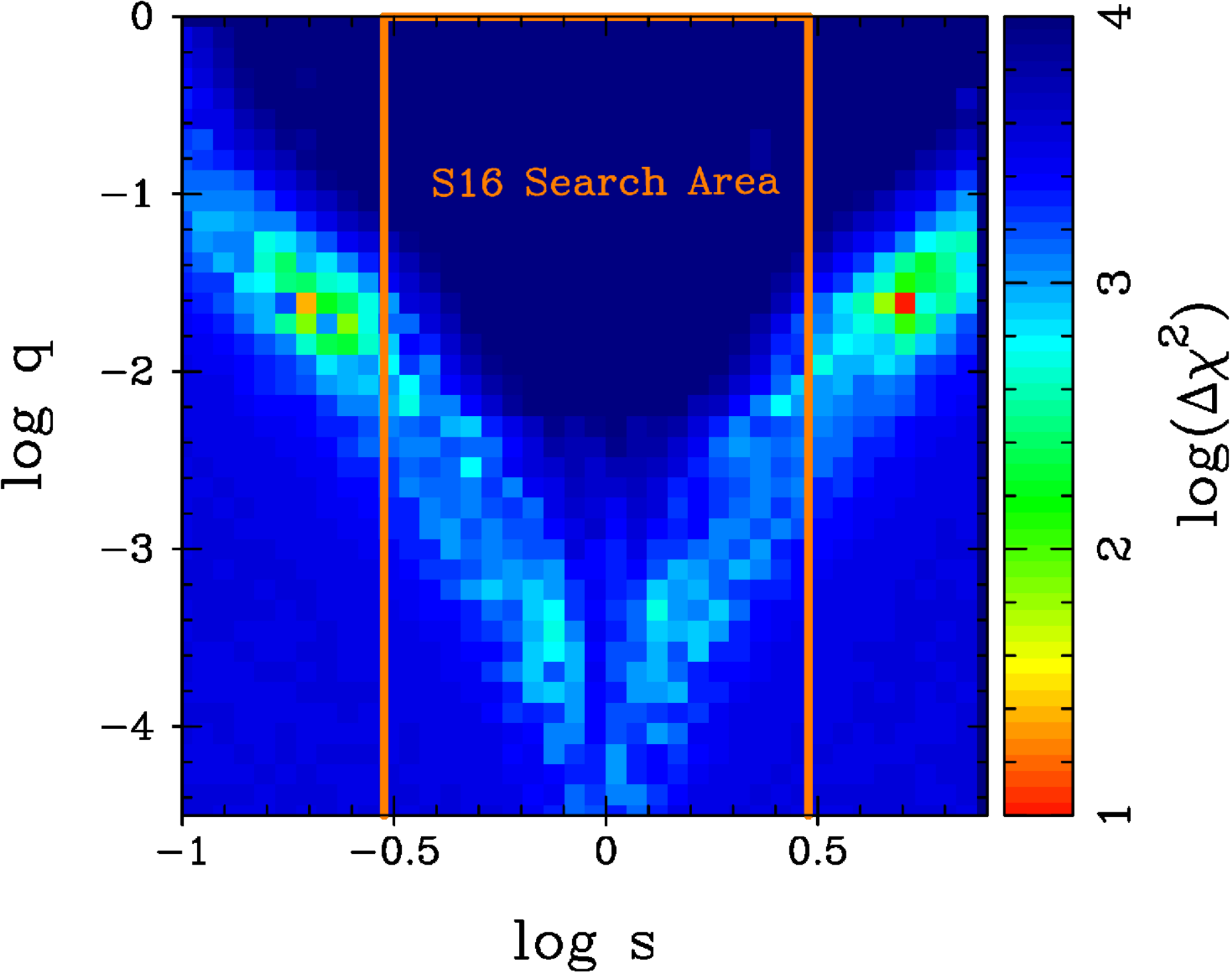}
\caption{
The map of the minimum $\Delta\chi^{2}$ in each $s$-$q$ grid from the grid search.
The orange box corresponds to the area of the grid search analysis in \citet{Shvartzvald+2016}.
}
\label{fig:chi2_map}
\end{figure}

At first, we explored the 2L1S interpretation to explain the anomalous features of the light curve. 
In modeling 2L1S microlensing light curves, it is common to encounter situations where different physical models explain the observed data equally well, e.g. the close/wide degeneracy \citep{Griest+1998,Dominik1999} and the planet/binary degeneracy \citep{Choi+2012,Miyazaki+2018}, where different combinations of the microlensing parameters can generate  morphologically similar light curves.  
Therefore, we should thoroughly investigate the multi-dimensional parameter space to find the global preferred model solution.
We conducted a detailed grid search over the $(q,s,\alpha)$ parameter space where the magnification pattern strongly depends on these three parameters.
The search ranges of $q$, $s$ and $\alpha$ are $-1<{\rm log}s<1$, $-4.5<{\rm log}q<0$ and $0<\alpha<2\pi$ with 40 grid points, respectively, and thus the total number of grid points is $40\times40\times40=64000$. 
We conducted the grid search analysis following the same procedure written in \citet{Miyazaki+2018}.
Figure \ref{fig:chi2_map} shows the map of the minimum $\Delta\chi^{2}$ in each $s$-$q$ grid from the grid search.
In Figure \ref{fig:chi2_map}, we found two possible local minima around $({\rm log}q,{\rm log}s)\sim(-1.8, -0.75)$ and $\sim(-1.8, 0.75)$, which is caused by the close/wide degeneracy.
After refining all the possible solutions, we found the best-fit 2L1S close ($s<1$) and wide ($s>1$) models with $q\sim0.03$, where the $\chi^{2}$ difference between them is only $\Delta\chi^{2}=4.9$.
As seen in Figure \ref{fig:LC_MB13551}, the 2L1S model with $q\sim0.03$ provide good fits to the anomalous features around the top of the light curve.
We also show the model light curve of 2L1S with $q\sim10^{-4}$ in Figure \ref{fig:LC_MB13551} and it does not fit the light curve anomaly well.
\vspace{0.1in}

S16 included this event in their statistical analysis as a planetary microlensing events, using a mass ratio of $q\sim10^{-4}$ for this event.
However, our reanalysis found that the static 2L1S models with $q\sim10^{-2}$ are preferred over the model with $q\sim10^{-4}$ by $\Delta\chi^{2}>700$.
The reason of the oversight is that models with $q\sim10^{-2}$ are outside of the range of their grid search of $-6<\log q<0$ and $0.3 < s < 3$.
And the search for the best-fit model outside of this range by refining model parameters found by their grid search was not conducted.
Another difference from S16 is that we used re-reduced MOA and OGLE light curves and included all the follow-up datasets.
However, we confirmed that the 2L1S models with $q\sim10^{-4}$ are disfavored relative to the models with $q\sim10^{-2}$ by $\Delta\chi^2>300$ even if we used the survey data, MOA, OGLE and Wise1m.
Therefore, note that the survey data were sufficient to identify the new solutions.

\subsubsection{Parallax Effects}
Although the best-fit static models provide good fits to the main anomaly features around the peak of the light curve, we found that, overall, the light curve slightly deviates from the static models.
The event OGLE-2013-BLG-0911 has $t_{\rm E}\sim90$ days and had continued throughout the bulk of the bulge season, which implies that the light curve could be affected by additional high-order microlensing effects.
\vspace{0.1in}

It is known that the orbital acceleration of Earth causes a parallax effect \citep{Gould1992,Gould2004,Smith+2003}.
This can be described by the microlensing parallax vector $\bm{\pi}_{\rm E}=(\pi_{{\rm E},N},\pi_{{\rm E},E})$.
Here, $\pi_{{\rm E},N}$ and $\pi_{{\rm E},E}$ denote the north and east components of $\bm{\pi}_{\rm E}$ projected to the sky plane in equatorial coordinates.
The direction of $\bm{\pi}_{\rm E}$ is defined so as to be identical to that of $\bm{\mu}_{\rm rel,G}$, which is the geocentric lens-source relative proper motion projected to the sky plane at a reference time $t_{\rm fix}$, and the amplitude of $\bm{\pi}_{\rm E}$ is $\pi_{\rm E}={\rm au}/\tilde{r}_{\rm E}$ where $\tilde{r}_{\rm E}$ is the Einstein radius projected inversely to the observer plane.
We took a reference time $t_{\rm fix}=6537.3$ days for this event.
The measurement of $\bm{\pi}_{\rm E}$ enables constraints to be placed on the relation between the lens mass $M_{L}$ and distance $D_{L}$ \citep{Gould2000,Bennett2008}.
For Galactic bulge source events, models with $(u_{0},\alpha,\pi_{{\rm E},N})$ and $-(u_{0},\alpha,\pi_{{\rm E},N})$ can yield very similar light curves \citep{Skowron+2011}.
This is reflected as a pair of the symmetric source trajectories to the binary and is sometimes referred to as ``ecliptic degeneracy''.
\vspace{0.1in}

Taking the parallax effect into consideration for modeling, we found that the two parallax parameters gave an improvement of $\Delta\chi^2\sim210$ compared to the best-fit static model. 
However, we also found that the best-fit parallax model seemed not to explain the long-term deviations of the light curve from the best-fit static model, as can be seen in Figure \ref{fig:LC_comb}.
This implies that there might still be other high-order microlensing effects in the light curve.
Note that adding the lens orbital motion does not improve our models.

\subsubsection{Xallarap Effects\label{sec:xallarap}}
\begin{figure}
\centering
\includegraphics[angle=-90, scale=0.6]{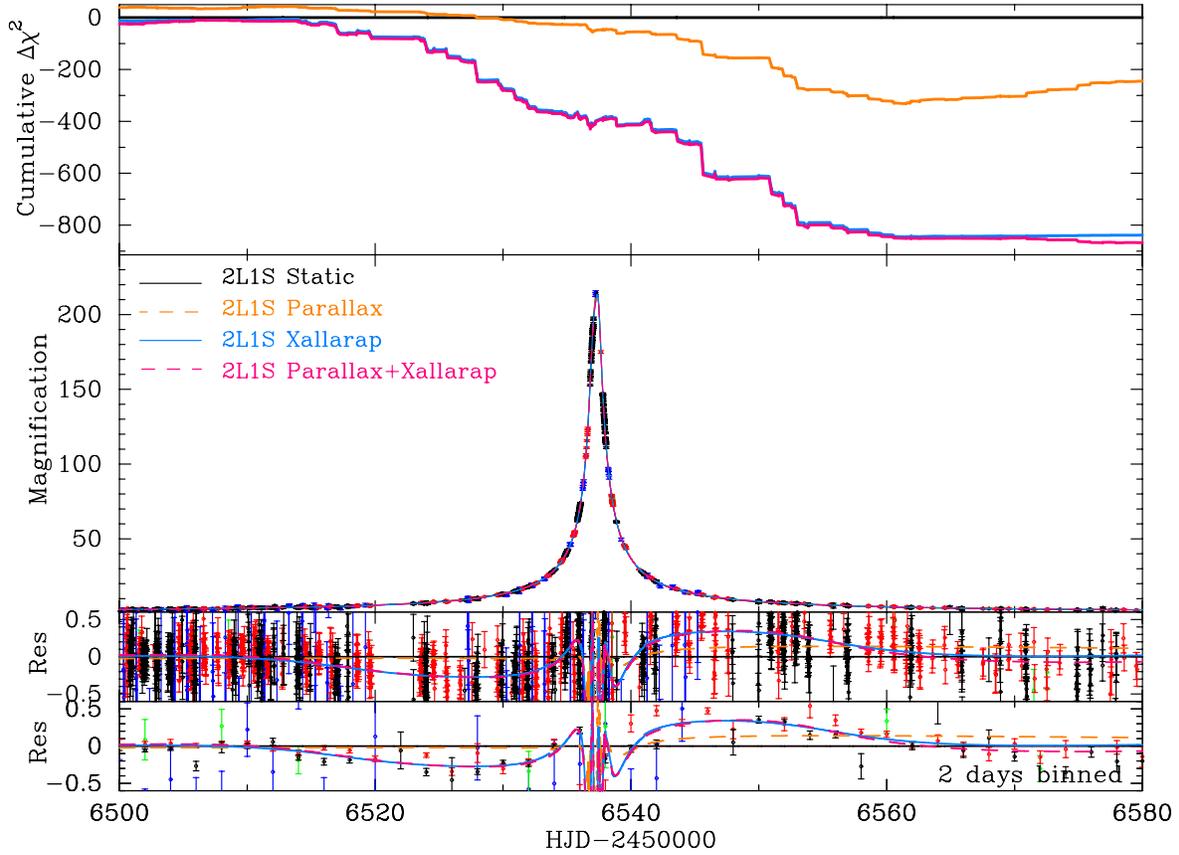}
\caption{
(Top) Cumulative $\Delta\chi^{2}$ distributions of the three 2L1S close ($u_{0}>0$) models compared to the 2L1S static model. 
(Second from the top) The light curve and models for OGLE-2013-BLG-0911. Here, we plot only MOA, OGLE and Wise1m light curves for clarity.
(Third from the top) The residuals of the light curve and models from the static model.
(Bottom) The residuals binned by 2 days.
}
\label{fig:LC_comb}
\end{figure}

Xallarap \citep{Griest+1992,Han+1997,Poindexter+2005} is the microlensing effect on the light curve induced by the source orbital motion around the source companion.
The xallarap model requires 7 additional fitting parameters which determine the orbital elements of the source system; the direction toward the solar system relative to the orbital plane of the source system, ${\rm R.A.}_{\xi}$ and ${\rm Dec.}_{\xi}$; the source orbital period, $P_{\xi}$; the source orbital eccentricity and perihelion time, $e_{\xi}$ and $T_{\rm peri}$; the xallarap vector, $\bm{\xi}_{\rm E}=(\xi_{{\rm E},N},\xi_{{\rm E},E})$.
The direction of $\bm{\xi}_{\rm E}$ is similar to that of the geocentric lens-source proper motion $\bm{\mu}_{\rm rel,G}$ and the amplitude of $\bm{\xi}_{\rm E}$ is $\xi_{\rm E}=a_{S}/\hat{r}_{\rm E}$ where $a_{S}$ is the semi-major axis of the source orbit and $\hat{r}_{\rm E}$ is the projected Einstein radius to the source plane, i.e., $\hat{r}_{\rm E}=\theta_{\rm E}D_{S}$.
Kepler's third and Newton's third laws give the following relations \citep{Batista+2009},
\begin{eqnarray}\label{eq:xallarap}
\xi_{\rm E} =  \frac{\rm 1\;au}{D_{S}\theta_{\rm E}}\left(\frac{M_{c}}{M_{\odot}}\right)\biggl[\frac{M_{\odot}}{M_{c}+M_{S}}\frac{P_{\xi}}{\rm 1\;year}\biggr]^{2/3}\ , \\
M_S a_S = M_C a_C \Rightarrow a_{SC} \equiv a_S + a_C = (1+\frac{M_S}{M_C})a_S\ ,
\end{eqnarray}
where $M_{S}$ and $M_{C}$ are the masses of the source and source companion, respectively.
Therefore, we can estimate the source companion mass $M_{C}$ from the xallarap measurements by assuming $M_{S}$ and $D_S$.
\vspace{0.1in}

Since the number of additional parameters for the xallarap effect is large, we conducted a grid search fixing $({\rm R.A.}_{\xi}, {\rm Dec.}_{\xi}, P_{\xi})$ in order to avoid missing any local minima.
After refining all the possible solutions, we found the best-fit xallarap model is favored over the best-fit parallax model by $\Delta\chi^{2}>650$.
As shown in Figure \ref{fig:LC_comb}, including the xallarap effect produces a model that fits the long-term residuals from the best-fit static model, and it dramatically improves the $\chi^{2}$ values.
The best-fit orbital period of the source system is $P_{\xi}\sim40$ days and is clearly different from Earth's orbital period of 365 days, which implies that the parallax and xallarap signals are clearly distinguishable.
Following Equation (\ref{eq:xallarap}), the best-fit 2L1S xallarap model indicates a source companion mass of  $M_{C}=0.21M_{\odot}$ and a distance between two sources $a_{SC}=0.22$ au on the assumption of $M_{S}=1.0M_{\odot}$ and $D_S=8$ kpc, which is a common stellar binary system in solar neighborhood \citep{Duchene+2013}. 
The best-fit $\xi_{\rm E}$ values are much smaller than 1, which means that the two sources are separated by much less than the Einstein radius.
Hence, the source companion was also likely to be magnified during the event.
In following sections, we explore the binary source scenarios where both components of the binary source system are magnified by the lens.

\subsection{Binary Source (1L2S) Model \label{sec:1L2S}}
When two source stars are magnified by the same single-lens, called a single-lens binary-source (1L2S) event, the observed flux would be the superposition of the  two magnified single-source fluxes, i.e.
\begin{equation}\label{eq:BS}
A(t)=\frac{A_{1}(t)F_{s,1}+A_{2}(t)F_{s,2}}{F_{s,1}+F_{s,2}}=\frac{A_{1}(t)+q_{F,j}A_{2}(t)}{1+q_{F,j}}\ ,
\end{equation}
where $A_{i}$ and $F_{s,i}$ represent the magnification and the baseline flux of each $i$-th source, and $q_{F,j}=F_{s,2}/F_{s,1}$ is the flux ratio between the two source stars in each $j$-th pass band.
For a standard (static) 1L2S model, the fitting parameters are [$t_{0}$, $t_{0,2}$, $t_{\rm E}$, $u_{0}$, $u_{0,2}$, $\rho$, $\rho_{2}$, $q_{F, j}$].
Because the magnification of each source star varies independently, the total observed source color is variable during a binary source event, which happens in single-source events only if limb-darkening  effects are seen during caustic crossings\footnote{For point lenses, this happens only if the lens briefly transits the source \citep{Loeb+1995,Gould+1996}} as  microlensing does not depend on wavelength.
Binary source events can mimic short-term binary-lens anomalies in a light curve, therefore it is necessary whether the anomaly features are induced by binary-lens or binary-source \citep{Gaudi1998,Jung+2017a, Jung+2017b, Shin+2019}.
\vspace{0.1in}

First, we fitted the light curves with the static 1L2S model and found that it was disfavored over the static 2L1S models by $\Delta\chi^{2}>1100$.
In Section \ref{sec:xallarap}, we found an asymmetric distortion in the light curves which can be explained by the xallarap effect (i.e. source orbital effect).
Thus, we also explored 1L2S models with source orbital motion. 
The trajectories of two sources can be estimated by the source orbital motion from the xallarap parameters, $(\xi_{{\rm E},N},\xi_{{\rm E},E},{\rm R.A.}_\xi, {\rm Dec.}_{\xi}, P_\xi, e_\xi, T_{\rm peri})$, and Equation (\ref{eq:xallarap}).
Here, we assumed $M_{S}=1$ $M_{\odot}$ and $D_S=8$ kpc to derive the source companion mass $M_C$.
In Appendix \ref{ap:msds}, we confirmed that the assumptions of $M_S=1\;M_{\odot}$ and $D_S=8\;{\rm kpc}$ hardly impact on the light curve modeling.
We conducted detailed grid search of $({\rm R.A.}_{\xi}, {\rm Dec.}_{\xi}, P_{\xi})$ and refined all the possible 1L2S solutions.
We found the best-fit 1L2S model is not preferred over the static 2L1S models by $\Delta\chi^2>80$ even if we introduced the source orbital motion.

\subsection{Binary-Lens Binary-Source (2L2S) Model}
\begin{figure}
\centering
\includegraphics[angle=-90, scale=0.55]{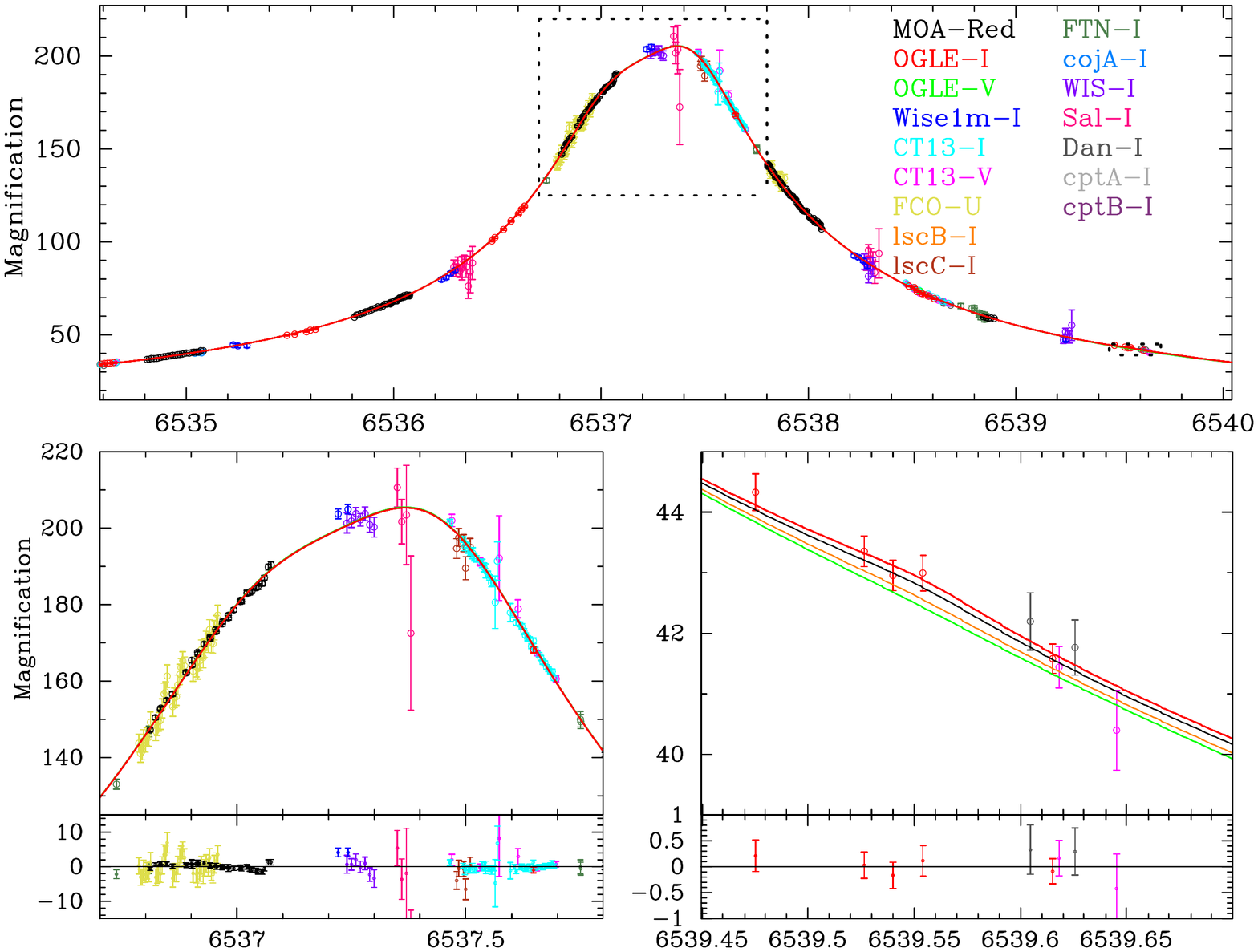}
\caption{
The light curve of OGLE-2013-BLG-0911.
Each color on the data point corresponds to each instrument, shown on the right.
The error bars are renormalized following Equation (\ref{eq:err}).
The 2L2S ($s<1$, $u_0>0$) model light curves in MOA-Red, $I$, $V$, Unfilered bands are shown as the solid black, red, green and orange lines, respectively.
The dotted boxes in the top panel correspond to the areas represented in the bottom left and right panels, where the primary and secondary sources were significantly magnified, respectively.
}
\label{fig:LC3}
\end{figure}

\begin{figure}
\centering
\includegraphics[angle=-90, scale=0.45]{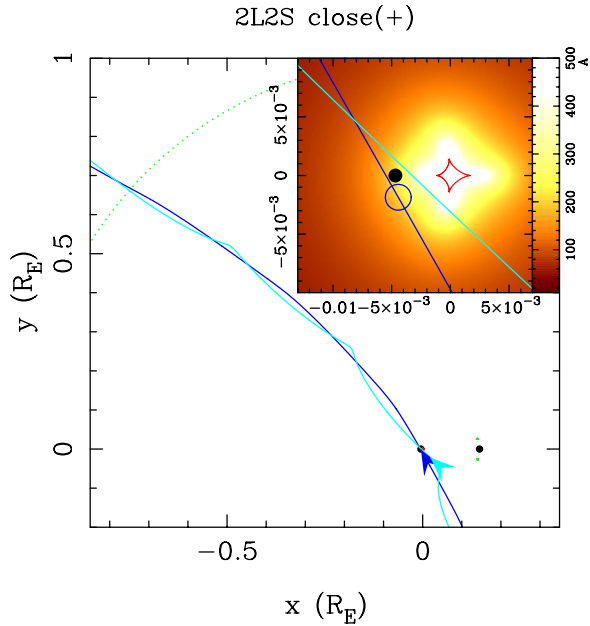}
\caption{
Caustic geometry for the best-fit 2L2S ($s<1, u_0>0$) model is shown as the red curves, respectively. 
The blue and light blue curves show the primary and secondary source trajectories with respect to the lens systems, with the arrows indicating the directions of each source motion.
The black dots are lens components and the green dots represent critical curves. 
The inset shows a zoom-in view around the central caustic.
The magnification patterns are described as color maps. 
The brighter tone denotes higher magnification.  
The blue circle on the lines indicates the primary source size and its positions is at $t_0$. 
}
\label{fig:caustic}
\end{figure}

\begin{deluxetable}{lccccccccccccc}
\centering
\tablecaption{The 2L2S Model Parameters}
\tablehead{
\multirow{2}{*}{Parameters}    & \multirow{2}{*}{Units} & \multicolumn{2}{c}{Close $(s<1)$} & \multicolumn{2}{c}{Wide $(s>1)$}  \\
 & & \multicolumn{1}{c}{($u_{0}>0$)} & \multicolumn{1}{c}{($u_{0}<0$)} & \multicolumn{1}{c}{($u_{c}>0$)} & \multicolumn{1}{c}{($u_{c}<0$)}    
}
\startdata
$t_{0}\;(t_c)$& HJD-2456530             & $7.3128^{+0.0005}_{-0.0005}$  & $7.3127^{+0.0005}_{-0.0005}$  & $7.3123^{+0.0003}_{-0.0004}$ & $7.3111^{+0.0005}_{-0.0006}$ \\
$t_{\rm E}$ & day                 & $94.698^{+1.612}_{-1.525}$    & $98.121^{+0.858}_{-0.958}$    & $101.104^{+2.246}_{-1.799}$    & $98.275^{+1.154}_{-1.148}$ \\
$u_{0}\;(u_c)$ & $(10^{-3})$             & $4.800^{+0.077}_{-0.079}$     & $-4.620^{+0.041}_{-0.042}$    & $4.522^{+0.086}_{-0.108}$      & $-4.626^{+0.053}_{-0.053}$ \\
$q$ & $(10^{-2})$                 & $3.236^{+0.089}_{-0.084}$     & $3.066^{+0.090}_{-0.094}$     & $3.160^{+0.132}_{-0.134}$      & $3.456^{+0.087}_{-0.067}$ \\
$s$ &                             & $0.150^{+0.002}_{-0.002}$     & $0.150^{+0.002}_{-0.002}$     & $6.774^{+0.101}_{-0.085}$      & $7.084^{+0.074}_{-0.064}$ \\
$\alpha$ & radian                 & $4.197^{+0.004}_{-0.004}$     & $2.078^{+0.006}_{-0.008}$     & $4.198^{+0.007}_{-0.007}$      & $2.092^{+0.006}_{-0.005}$ \\
$\rho$ & $(10^{-3})$              & $1.113^{+0.148}_{-0.118}$     & $1.136^{+0.090}_{-0.144}$     & $1.413^{+0.093}_{-0.295}$      & $0.971^{+0.118}_{-0.085}$ \\
$\pi_{{\rm E},N}$ &               & $0.256^{+0.044}_{-0.050}$     & $0.300^{+0.027}_{-0.025}$     & $0.319^{+0.039}_{-0.050}$      & $0.271^{+0.032}_{-0.041}$ \\
$\pi_{{\rm E},E}$ &               & $0.018^{+0.005}_{-0.005}$     & $0.004^{+0.006}_{-0.006}$     & $0.001^{+0.006}_{-0.006}$      & $0.006^{+0.003}_{-0.004}$ \\
$\xi_{{\rm E},N}$ & $(10^{-3})$   & $-2.91^{+0.97}_{-1.01}$    & $-3.13^{+1.72}_{-1.39}$    & $-5.32^{+2.66}_{-1.61}$     & $-2.48^{+1.28}_{-1.04}$ \\
$\xi_{{\rm E},E}$ & $(10^{-3})$   & $-4.31^{+0.18}_{-0.17}$    & $-3.59^{+0.62}_{-0.44}$    & $-3.53^{+0.45}_{-0.48}$     & $-3.66^{+0.50}_{-0.33}$ \\
${\rm R.A.}_{\xi}$ & degree       & $-74.2^{+12.3}_{-12.2}$ & $-87.4^{+15.5}_{-14.0}$ & $260.9^{+13.5}_{-11.4}$  & $-89.4^{+16.3}_{-16.3}$ \\
${\rm Dec.}_{\xi}$ & degree       & $21.8^{+6.4}_{-7.4}$    & $29.8^{+2.5}_{-3.5}$    & $19.7^{+2.2}_{-6.2}$     & $38.0^{+7.4}_{-7.7}$ \\
$P_{\xi}$ & day                   & $36.67^{+0.77}_{-0.73}$    & $36.28^{+0.74}_{-0.70}$    & $36.82^{+0.66}_{-0.68}$     & $36.51^{+0.80}_{-0.70}$ \\
$e_{\xi}$ &                       & $0.258^{+0.033}_{-0.029}$     & $0.249^{+0.029}_{-0.031}$     & $0.270^{+0.032}_{-0.029}$      & $0.231^{+0.040}_{-0.038}$ \\
$T_{\rm peri}$ & HJD-2456500      & $53.14^{+1.08}_{-1.10}$  & $52.75^{+0.48}_{-0.51}$  & $17.31^{+0.90}_{-0.84}$   & $53.69^{+0.78}_{-0.96}$ \\
$q_{F,Red}$ & $(10^{-3})$         & $1.122^{+0.388}_{-0.330}$     & $0.818^{+0.317}_{-0.254}$     & $0.974^{+0.439}_{-0.298}$      & $0.970^{+0.342}_{-0.261}$ \\
$q_{F,I}$ & $(10^{-3})$           & $1.424^{+0.466}_{-0.399}$     & $1.058^{+0.383}_{-0.309}$     & $1.246^{+0.527}_{-0.362}$      & $1.241^{+0.412}_{-0.317}$ \\
$q_{F,V}$ & $(10^{-4})$           & $3.58^{+1.58}_{-1.29}$     & $2.39^{+1.24}_{-0.93}$     & $3.00^{+1.76}_{-1.12}$      & $2.98^{+1.36}_{-0.99}$ \\
$q_{F,R}$ & $(10^{-4})$           & $6.62^{+2.64}_{-2.18}$     & $4.61^{+2.10}_{-1.62}$     & $5.63^{+2.96}_{-1.93}$      & $5.60^{+2.30}_{-1.70}$ \\
$\pi_{\rm E}$ &                   & $0.257^{+0.044}_{-0.050}$     & $0.300^{+0.027}_{-0.025}$     & $0.319^{+0.039}_{-0.050}$      & $0.271^{+0.032}_{-0.041}$ \\
\hline\hline
$\chi^{2}$ & & \multicolumn{1}{c}{16542.2} & \multicolumn{1}{c}{16543.3} & \multicolumn{1}{c}{16542.3} & \multicolumn{1}{c}{16542.9} \\
$\Delta\chi^{2}$ & & \multicolumn{1}{c}{-} & \multicolumn{1}{c}{1.1} & \multicolumn{1}{c}{0.1} & \multicolumn{1}{c}{0.7} 
\enddata
\label{tab:2L2S}
\tablecomments{
Here, we assume $M_S=1\;M_{\odot}$ and $D_S=8\;{\rm kpc}$.
The flux ratios $q_{F}$ and parallax amplitude $\pi_{\rm E}=\sqrt{\pi^2_{\rm E,N}+\pi^2_{\rm E,E}}$ are not fitting parameters.
All the other parameters in this table are used as fitting parameters for modeling. 
}
\end{deluxetable}

Finally, we explored the 2L2S models with source orbital motion, i.e., taking account the flux from the source companion and the xallarap effect.
Here, we adopted the flux ratios $q_F$ estimated from $M_C$ which is derived from the xallarap parameters to keep the consistency.
We derived the flux ratios in each band from a combination of $M_C$ and a theoretical stellar isochrone model\footnote{http://stev.oapd.inaf.it/cgi-bin/cmd} (PARSEC; \citealp{Bressan+2012}) for solar metallicity and a typical bulge star age of 10 Gyr.
For the MOA-Red band, we derived the flux ratio from that in $I$- and $V$- bands, $q_{F,Red}=q_{F,I}^{0.827}q_{F,V}^{0.173}$.
This formula comes from the following color transformation that is derived by using bright stars around the event \citep{Gould+2010,Bennett+2012,Bennett+2018},
\begin{equation}
R_{\rm MOA} - I_{\rm O3} = 0.173(V_{\rm O3}-I_{\rm O3})+{\rm const}
\end{equation}
where $R_{\rm MOA}$, $I_{\rm O3}$ and $V_{\rm O3}$ are the magnitudes in MOA-Red, OGLE-III $I$- and $V$-bands, respectively.
For the Unfiltered passband, we used the $R$-band flux ratio assuming $q_{F, {\rm Unfiltered}}\approx q_{F, R}$. 
\vspace{0.1in}

We found the four best 2L2S models, which suffer from  the close/wide degeneracy and the ecliptic degeneracy. 
The parameters of these models are shown in Table \ref{tab:2L2S}.
The light curve of the best-fit 2L2S $(s<1$, $u_0>0)$ model  is shown in Figure \ref{fig:LC3}.
Here, as shown in Equation (\ref{eq:BS}), the light curves in each passband are different. 
The black, red, green and cyan solid curves indicate the model light curves in the MOA-Red, $I$-, $V$- and $R$-bands, respectively.
The caustic geometry and source trajectories of the best-fit 2L2S $(s<1$, $u_0>0)$ model are shown in Figure \ref{fig:caustic}.
Here the source companion trajectory indicates that the source companion is more strongly magnified than the primary source.
In general, such magnification differences in two sources, allow us to resolve the close/wide degeneracy and the ecliptic degeneracy. 
However, as shown in the bottom right panel of Figure \ref{fig:LC3}, where the secondary source magnification is peaked at HJD'$\sim6539.55$ , the flux contribution is $\sim0.01$ times smaller than the primary source because the source companion is intrinsically much fainter than the primary source. 
Consequently, we could not resolve these degeneracies.
These 2L2S models are preferred relative to the 2L1S models with parallax and xallarap effects by $\Delta\chi^2\sim16$
 without additional fitting parameters.
The fitting and physical parameters for the 2L1S and 2L2S models are almost identical each other.
Therefore, it hardly affects the final results whichever we take.
Hereafter, we take the 2L2S models for the final result. 

\section{Source Properties\label{sec:source}}
The measurement of $\rho$ enables us to determine the angular Einstein radius $\theta_{\rm E}=\theta_{*}/\rho$ where $\theta_{*}$ is the angular source radius.
The angular source radius $\theta_{*}$ can be estimated from the extinction-free color and magnitude of the source star by using a method similar to that of \citet{Yoo+2004} which adopts the centroid of the bulge red clump giants (RCG) as a reference point.
\citet{Yoo+2004} assumed that the source star suffers from the same extinction as that of the bulge RCG so that the extinction-free color and magnitude of the source star can be described as the following equation, 
\begin{equation}
(V-I, I)_{S,0}= (V-I, I)_{0,{\rm RCG}}-\Delta(V-I, I)\ ,
\end{equation}
where  $(V-I, I)_{0,{\rm RCG}}=(1.06\pm0.07, 14.40\pm 0.04)$ is the extinction-free color and magnitude of the bulge RCG centroid \citep{Bensby+2011, Bensby+2013, Nataf+2013} and $\Delta(V-I, I)$ are the offsets of the color and magnitude from the RCG centroid to the source star measured in the standard color magnitude diagram (CMD).

\subsection{Photometric Source Properties}

\begin{figure}[htb!]
\centering
\includegraphics[angle=-90, scale=0.55,clip]{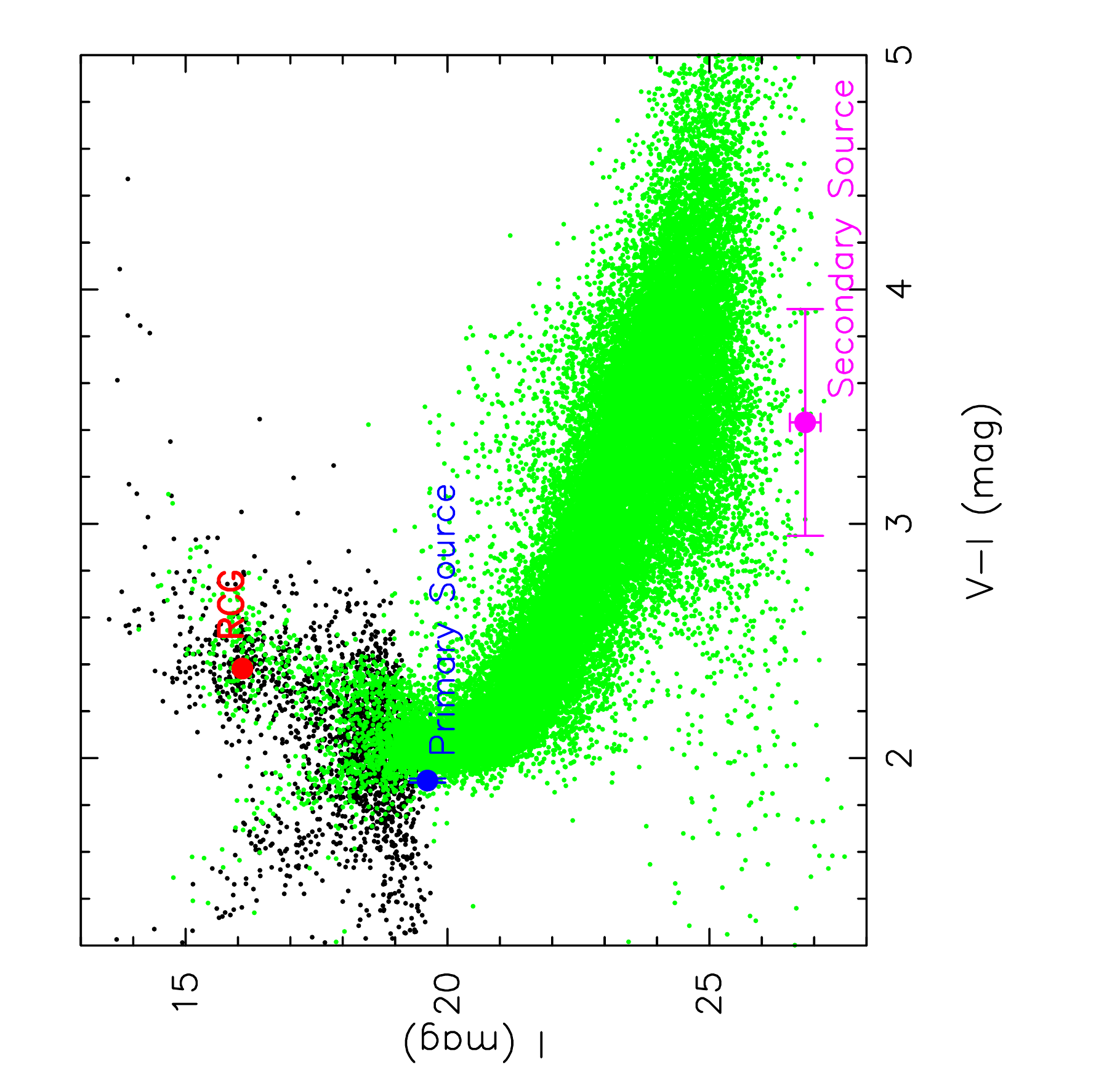}
\caption{
The $(V-I,I)$ color magnitude diagram (CMD) in the standard Kron-Cousins $I$ and Johnson $V$ photometric system. 
The positions of the primary and secondary source and the centroid of RCG are represented as the blue, magenta and red circles.  
The black dots indicate the OGLE-III catalog stars within $1^\prime$ of the source.
The green dots indicate the {\it Hubble Space Telescope} CMD in Baade's window \citep{Holtzman+1998} whose color and magnitude are matched by using the RCG position.
}
\label{fig:cmd}
\end{figure}

We obtained the apparent source color and magnitude of $(V-I, I)_{S}=(1.904\pm0.008, 19.618\pm0.006)$ derived from the measurements of CT13-$I$ and $V$ in the light curve modeling, which is detailed in Appendix \ref{ap:calibration}.
We also derived the source color and magnitude from the measurements of OGLE-$I$ and $V$ and confirmed that they are consistent within $2\sigma$, which is also detailed in Appendix \ref{ap:calibration}.
In addition, we independently measured the source color using a linear regression from CT13-$I$ and $V$, $(V-I)_{\rm CT13, reg}=1.910\pm0.005$, which is consistent with $(V-I)_{S}$.
Therefore, we judged the measurements of the source color and magnitude are robust.
Here, we took the source color and magnitude derived from the CT-13 measurements because both CT13-$I$ and -$V$ covered the light curve well when the primary source were significantly magnified. 
\vspace{0.1in}

Figure \ref{fig:cmd} shows the CMD of the OGLE-III catalog within $60^{''}$ of the sources plotted as black dots, and the CMD of Baade's window from \citet{Holtzman+1998} plotted as green dots. 
We found that the extinction-free color and magnitude of the primary source star are $(V-I, I)_{S,0}=(0.582\pm0.071, 17.936\pm0.049)$ assuming that the source suffers from the same extinction of the RCG centroid of $(E(V-I), A_{I})_{\rm RCG}=(1.322\pm0.071, 1.682\pm0.049)$.
The primary and secondary source stars are represented as blue and magenta dots in Figure \ref{fig:cmd}.
The primary source star seems to be somewhat bluer and brighter than other typical bulge dwarfs, which implies that the source possibly suffered less from reddening and extinction than the bulge RCG centroid.  

\subsection{Spectroscopic Source Properties}

\citet{Bensby+2017} took a spectrum of OGLE-2013-BLG-0911S and reported the source properties in detail, which are summarized in Table \ref{tab:source}\footnote{http://cdsarc.u-strasbg.fr/viz-bin/qcat?J/A+A/605/A89}. 
They suggested a possibility that the source star belongs to the foreground Galactic disk for three reasons. 
First, they measured the lens-source relative proper motion of $\mu\sim0.3$ mas/yr based on their single-lens microlensing model and indicated that this small value preferred the foreground disk source. 
Second, the intrinsic source color $(V-I)_{S,0}=0.71^{+0.03}_{-0.02}$ based on their spectroscopic measurement is redder than $(V-I)_{S,0}=0.49$ from their microlensing analysis, which implies that the source suffers less extinction than the average RCG in this field. 
They suggested that this may be because the source is in the foreground disk.
Note that our derived $(V-I)_{S,0}$ assuming the source is behind all of the dust, is less blue (0.58 vs. 0.49), but is still substantially bluer than Bensby's spectroscopic value. 
Third, they claimed that the heliocentric radial velocity of the source, ${\rm RV}_{\rm helio}$, is consistent with a disk star.
\vspace{0.1in}

However, if we adopt $I_{S,0} = 17.94$ which is derived from our light curve modeling and the absolute source magnitude $M_I=2.98$ which is estimated from spectroscopic values in \citet{Bensby+2017}, these measurements yield a source distance of $\sim9.8$ kpc, which would put the source within or behind the bulge. 
Furthermore, we consider that the above rationale for the disk source scenario is not strong for three reasons. 
First, both our 2L2S and 2L1S models provided $\mu\sim3\;{\rm mas/yr}$ which does not strongly favor the foreground disk source.
It is likely that their 1L1S model which could not fit the light curve properly, derived the incorrect values of $\mu\sim0.3\;{\rm mas/yr}$ and $(V-I)_{S,0}=0.49$.
Second, the color of $(V-I)_{S,0}=0.58\pm0.07$ derived in our analysis is between their spectroscopic and microlensing values and we found a similar color with their microlensing value when we use RCGs in slightly wider area around the target, where the RCG distribution gets spread wider along the extinction vector on the CMD.
These indicate that their spectroscopic color $(V-I)_{S,0}=0.71^{+0.03}_{-0.02}$ is correct and their and our photometric color $(V-I)_{S,0}=0.49$ and $0.58\pm0.07$, respectively, which are based on the average color of RCG in wider area, are biased because of low spatial resolution relative to the actual spacial variation of the reddening. 
Therefore, we conclude that the color difference may be due to the local spatial variation of the extinction in this field rather than the foreground disk scenario.
Third, the constraint from ${\rm RV}_{\rm helio}$ is not strong because it is also sufficiently explained by the bulge velocity distribution which has a large dispersion of $\sigma\sim100\;{\rm km/s}$ \citep{Howard+2008}.
\vspace{0.1in}

Finally, we adopt 90\% of the RCG extinction as the source extinction, i.e., $(E(V-I), A_I)_S=0.9\times(E(V-I),A_I)_{\rm RCG}=(1.190\pm0.064,1.514\pm0.044)$ and thus the intrinsic primary source color and magnitude are $(V-I, I)_{S,0}=(0.714\pm0.071, 18.109\pm0.049)$.
This is consistent with the spectroscopic source color $(V-I)_{S,0}=0.71^{+0.03}_{-0.02}$.
Note that even if we assumed that the source suffered from the same extinction as that for average RCG, the estimated source angular radius $\theta_{*}$ is consistent with that with 90\% of the average RCG extinction.
The source properties are summarized in Table \ref{tab:source}.

\begin{deluxetable}{lccc}[htb!]
\tablecaption{The Source Properties}
\tablehead{
 & $V-I$(mag) & $I$(mag) & $\theta_{*}$($\mu$as) 
}
\startdata
apparent & $1.904\pm0.009$ & $19.618\pm0.006$& - \\
intrinsic & $0.714\pm0.071$ & $18.104\pm0.049$ & $0.757\pm0.054$\\ \hline \hline 
& \multicolumn{2}{c}{From \citet{Bensby+2017}} & \\ \hline
Effective Temperature & $T_{\rm eff}$\tablenotemark{$a$} & \multicolumn{2}{c}{$5785\pm77$\;(K)} \\
 & $T_{\rm eff}$\tablenotemark{$b$} & \multicolumn{2}{c}{$6616$\;(K)} \\
Source Color & $(V-I)_{S,0}$\tablenotemark{$a$} & \multicolumn{2}{c}{$0.71^{+0.03}_{-0.02}$\;(mag)} \\
 & $(V-I)_{S,0}$\tablenotemark{$b$} & \multicolumn{2}{c}{0.49\;(mag)} \\
Absolute Magnitude & $M_{V}$\tablenotemark{$a$} & \multicolumn{2}{c}{3.69\;(mag)} \\
Heliocentric Radial Velocity & ${\rm RV}_{\rm helio}$ & \multicolumn{2}{c}{$-46.8$\;(km/s)}
\enddata
\tablecomments{\citet{Bensby+2017} modeled OGLE-2013-BLG-0911 as a 1L1S event.}
\tablenotetext{a}{Derived from spectroscopy.}
\tablenotetext{b}{Derived from their microlensing model.}
\label{tab:source}
\end{deluxetable}

\subsection{Angular Source and Einstein Radius}
With the extinction-free color and magnitude of the source, we can estimate $\theta_{*}$ from a precise empirical $(V-I)$ and $I$ relation 
\begin{equation}\label{eq:source_radius}
{\rm log_{10}}\left(\frac{2\theta_{*}}{{\rm mas}}\right)=0.5014+0.4197(V-I)_{S,0}-0.2I_{S,0}\ ,
\end{equation}
which is the optimized relation for the color ranges of microlensing observation, derived from the extended analysis of \citet{Boyajian+2014}.
Using Equation (\ref{eq:source_radius}), we estimated $\theta_{*}= 0.757\pm0.054$ $\mu$as for the best-fit model.
We used Equation (\ref{eq:source_radius}) and took account of the source extinction and its uncertainty into our MCMC calculations to derive the angular Einstein radius $\theta_{\rm E}$ and the geocentric lens-source relative proper motion $\mu_{\rm rel,G}$ for each model.  
The results are summarized in Table \ref{tab:property}.

\section{Lens System Properties\label{sec:lens_priority}}
\begin{figure}
\centering
\begin{minipage}[b]{0.45\linewidth}
\centering
\bf{\large Close Model}\hspace{6cm} 
\includegraphics[angle=-90, scale=0.35]{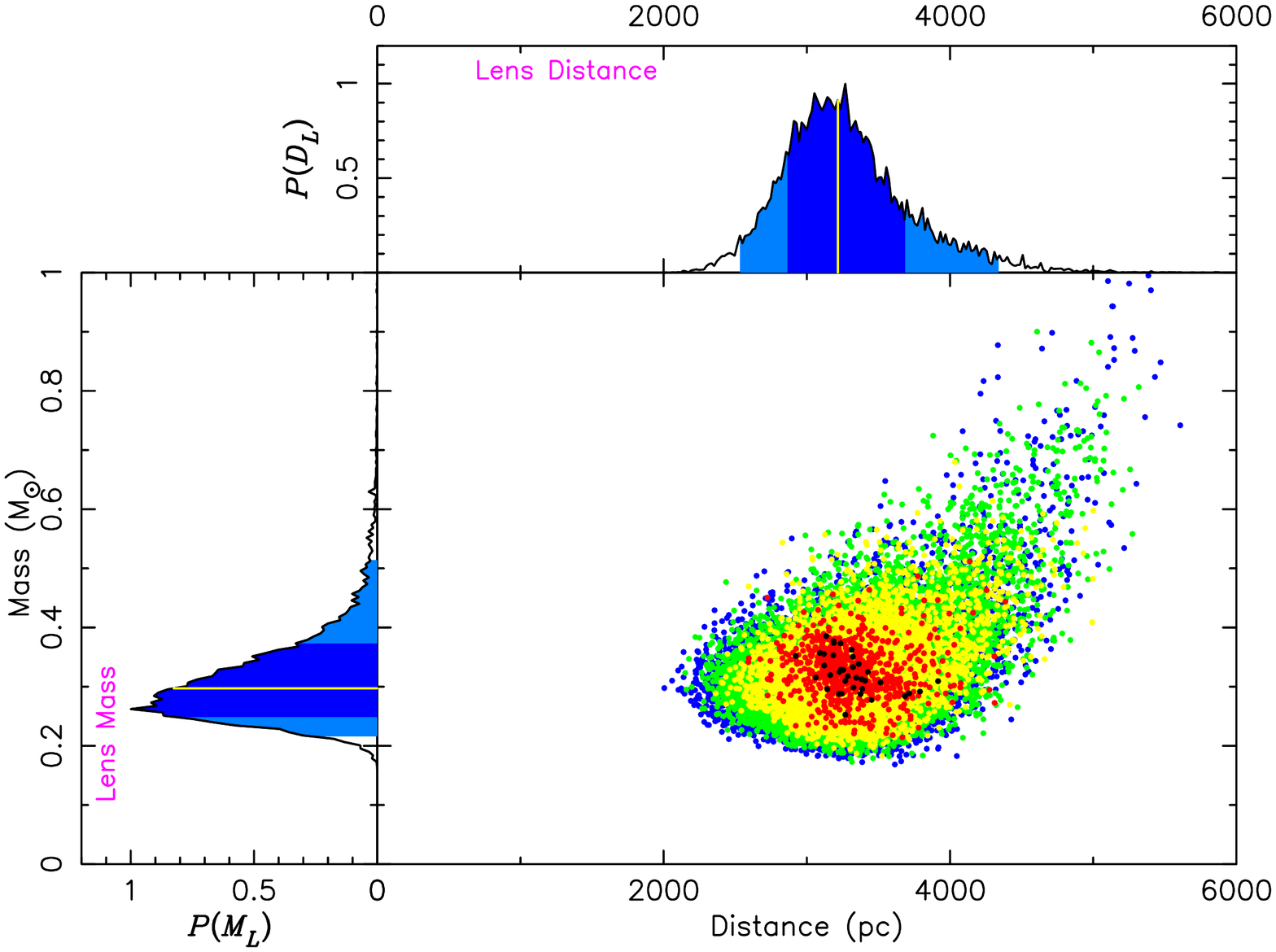} 
\end{minipage}
\begin{minipage}[b]{0.45\linewidth}
\centering
\bf{\large Wide Model}\hspace{6cm} 
\includegraphics[angle=-90, scale=0.35]{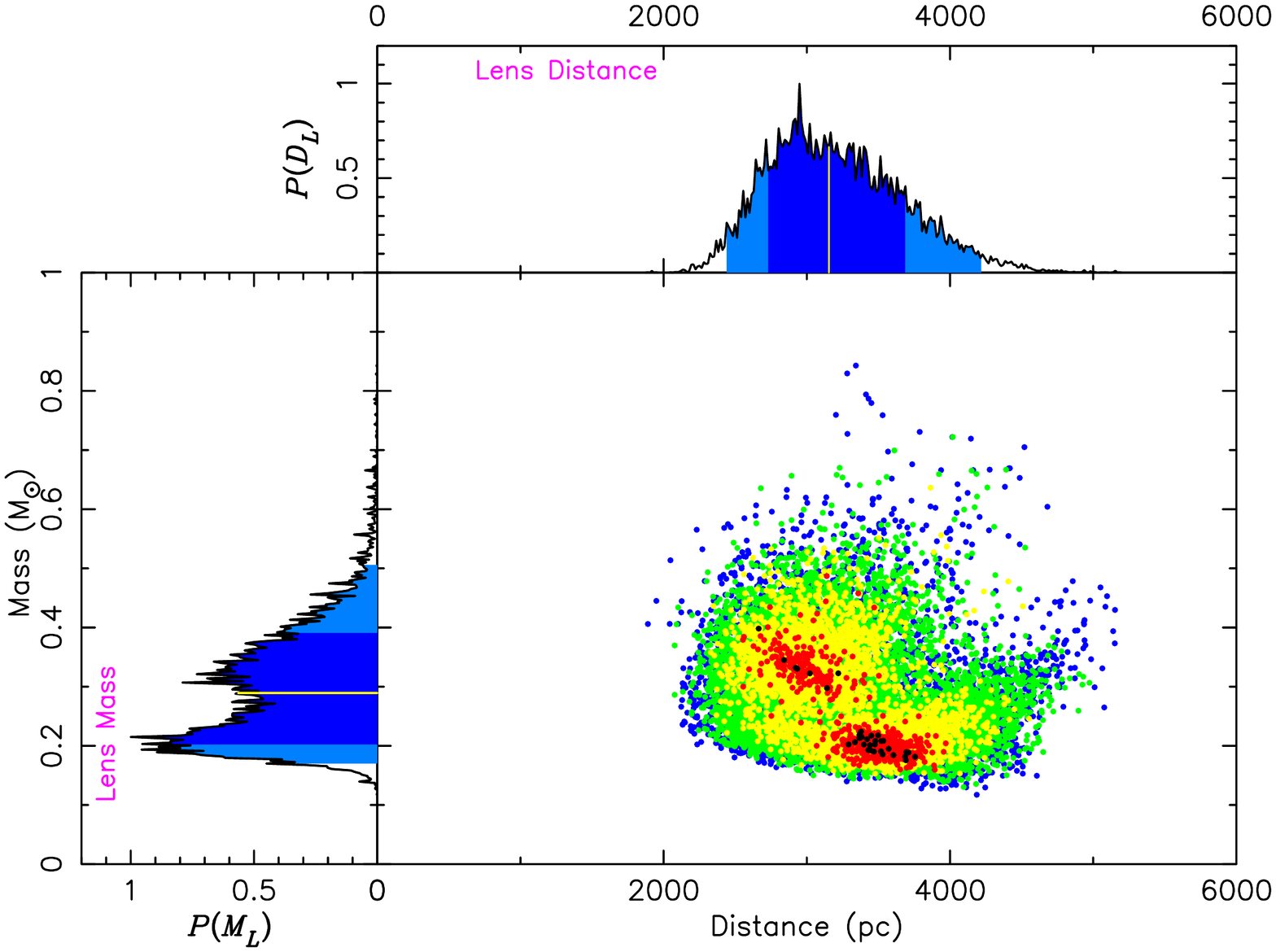} 
\end{minipage}
\caption{
The main panel shows $\Delta\chi^{2}$ distribution of the lens mass $M_{L}$ and distance $D_{L}$ for the close and wide models derived from MCMC, where black, red, yellow, green and blue dots indicate links with $\Delta\chi^{2}<1$, 4, 9, 16 and 25, respectively.
Top and left insets represent the posterior probability distributions of $M_{L}$ and $D_{L}$, where the dark and light blue regions indicate the 68.3\% and 95.4\% confidence interval, and the perpendicular yellow lines indicate the median values.
}
\label{fig:MLDL}
\end{figure}

\begin{deluxetable}{lcccccccccc}
\tablecaption{Physical Parameters}
\tablehead{
 \multicolumn{1}{c}{Parameters}    &  \multicolumn{1}{c}{Units} & \multicolumn{1}{c}{Close} & \multicolumn{1}{c}{Wide} 
}
\startdata
 Lens Host Mass, 			$M_{\rm host}$  & $M_{\odot}$      & $0.29^{+0.07}_{-0.05}$    & $0.28^{+0.10}_{-0.08}$ \\
Lens Companion Mass,   $M_{\rm comp}$ & $M_{\rm Jup}$    & $9.51^{+2.72}_{-1.69}$    & $9.92^{+3.78}_{-3.45}$ \\
Lens Distance,                 $D_L$ & kpc                  & $3.22^{+0.47}_{-0.35}$       & $3.15^{+0.53}_{-0.42}$ \\
Expected Semi-major Axis, $a_{\rm exp}\tablenotemark{1}$ & au                  & $0.39^{+0.05}_{-0.03}$    & $17.98^{+3.21}_{-3.24}$ \\
Source Companion Mass,  $M_{C}$ & $M_{\odot}$                     & $0.137^{+0.018}_{-0.016}$    & $0.137^{+0.017}_{-0.014}$ \\
Distance between Sources, $a_{SC}$ &             au            & $0.225^{+0.004}_{-0.004}$    & $0.225^{+0.003}_{-0.003}$ \\
Angular Einstein Radius,      $\theta_{\rm E}$ & mas            & $0.67^{+0.10}_{-0.08}$    & $0.68^{+0.14}_{-0.17}$ \\
Geocentric Lens-Source Proper Motion, $\mu_{\rm rel, G}$ & mas/yr          & $2.54^{+0.37}_{-0.30}$    & $2.50^{+0.56}_{-0.65}$ \\
Predicted Lens Magnitude,  $V_L$ & mag &  \multicolumn{2}{c}{$26.42_{-1.13}^{+1.15}$} \\
Predicted Lens Magnitude, $I_L$ & mag & \multicolumn{2}{c}{$22.80_{-0.83}^{+0.88}$}  \\
Predicted Lens Magnitude, $H_L$ & mag  & \multicolumn{2}{c}{$19.99_{-0.78}^{+0.79}$}  \\
Predicted Lens Magnitude, $K_L$ &  mag & \multicolumn{2}{c}{$19.64_{-0.76}^{+0.78}$} 
\enddata
\label{tab:property}
\tablecomments{The median value and 68.3\% confidence interval derived from MCMC. Here, we assume $D_S=8\;{\rm kpc}$ and $M_S=1M_{\odot}$ except for the lens magnitudes.}
\tablenotetext{1}{$a_{\rm exp}=\sqrt{3/2}a_{\perp}$.}
\end{deluxetable}

The measurements of both $\theta_{\rm E}$ and ${\pi}_{\rm E}$ enable us to determine the lens mass $M_{L}$ and distance $D_{L}$ directly \citep{Gould2000,Bennett2008}  as follows  
\begin{eqnarray}\label{eq:mass-distance}
M_{L} &=& \frac{c^2}{4G}\theta^2_{\rm E}\frac{D_{S}D_{L}}{D_S-D_L} = \frac{c^2}{4G}\frac{\rm au}{\pi^2_{\rm E}}\frac{D_S-D_L}{D_{S}D_{L}} =\frac{\theta_{\rm E}}{\kappa\pi_{\rm E}}\ ,
\end{eqnarray}
where $D_{L}$ is the lens distances.
We derived the probability distributions of the physical parameters of the source and lens systems by calculating their values in each MCMC link.
Here, we assumed the primary source mass $M_S=1M_{\odot}$ and the source distance $D_S=8$ kpc.
As referred in Appendix \ref{ap:msds}, we confirmed that the assumptions hardly affect the MCMC posterior distributions for the lens physical parameters except for the lens distance $D_L$.
We combined the posterior probability distributions of each model weighting by $e^{-\Delta\chi^2/2}$. 
Figure \ref{fig:MLDL} shows the probability distributions of the lens mass $M_{L}$ and distance $D_{L}$ for the close and wide models, and the final result of the physical parameters are summarized in Table \ref{tab:property}.
The result indicates that the lens system is an M-dwarf orbited by a massive Jupiter companion at very close ($M_{\rm host}=0.30^{+0.08}_{-0.06}M_{\odot}$, $M_{\rm comp}=10.1^{+2.9}_{-2.2}M_{\rm Jup}$, $a_{\rm exp}=0.40^{+0.05}_{-0.04}{\rm au}$) or wide ($M_{\rm host}=0.28^{+0.10}_{-0.08}M_{\odot}$, $M_{\rm comp}=9.9^{+3.8}_{-3.5}M_{\rm Jup}$, $a_{\rm exp}=18.0^{+3.2}_{-3.2}{\rm au}$) separation.
\vspace{0.1in}

\begin{figure}
\centering
\includegraphics[angle=-90, scale=0.4,clip]{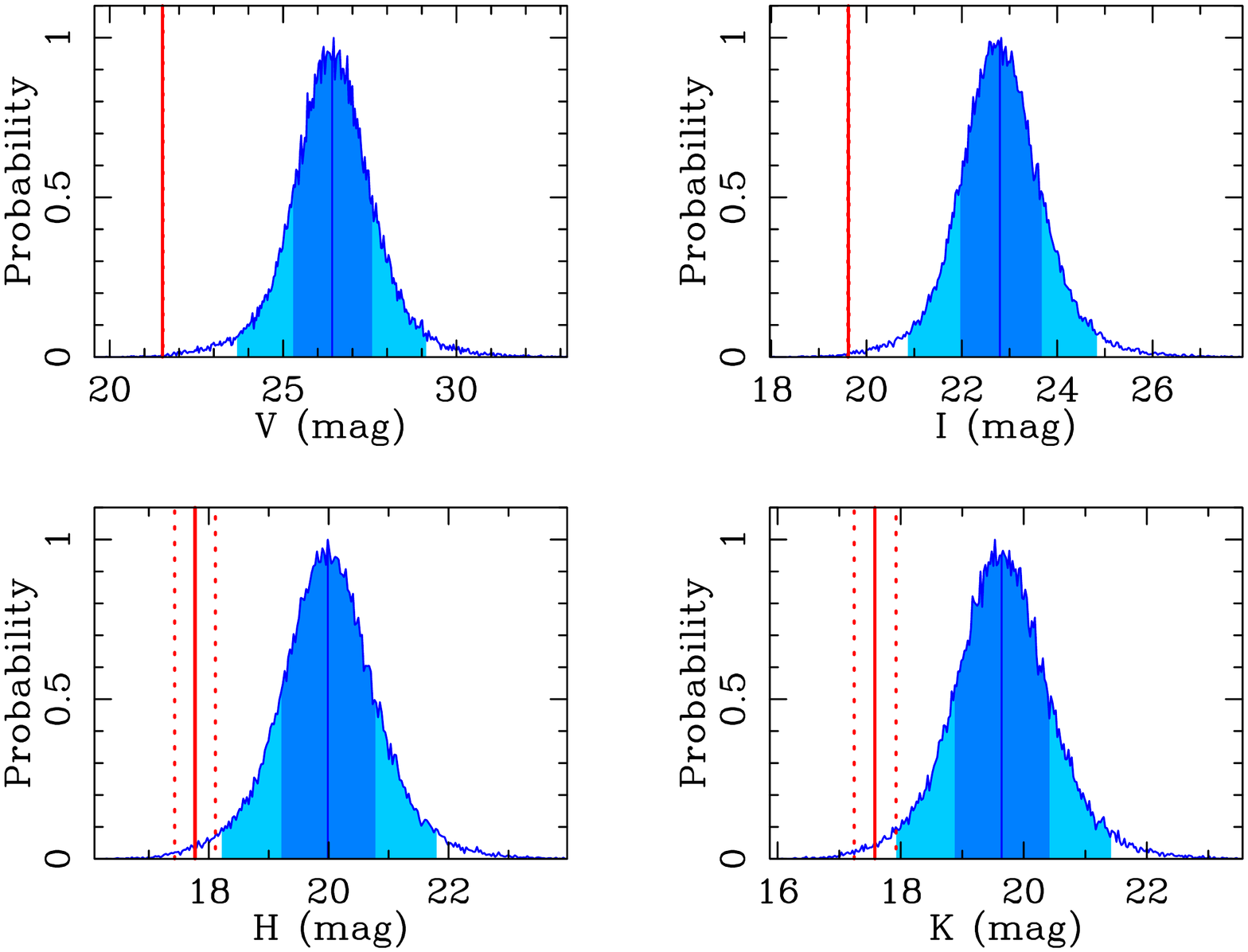}
\caption{
Posterior probabilities of the lens apparent magnitudes derived from the Bayesian analysis with the observed $t_{\rm E}$, $\theta_{\rm E}$ and $\pi_{\rm E}$ and prior probabilities from a standard Galactic model.
The dark and light blue regions indicate the 68.3\% and 95.4\% confidence interval, and the perpendicular blue lines indicate the median values.
The vertical solid and dashed red lines are the source magnitudes and its $1\sigma$ uncertainties in each passband.
}
\label{fig:lens_mag}
\end{figure}

We evaluated the expected apparent magnitude of the lens brightness by conducting a Bayesian analysis based on the observed $t_{\rm E}$, $\theta_{\rm E}$ and $\pi_{\rm E}$ and prior probabilities from a standard Galactic model \citep{Sumi+2011}.
Here, we evaluated the extinction in front of the lens given by 
\begin{equation}
A_{i, L} = \frac{1-e^{-D_L/h_{\rm dust}}}{1-e^{-D_S/h_{\rm dust}}}A_{i,S}\ ,
\end{equation} 
where the index $i$ corresponds to the passband $V$, $I$, $H$ and $K$, and the $h_{\rm dust}=(0.1 {\rm\;kpc})/{\rm sin}|b|$ is a scale length of the dust toward the event \citep{Bennett+2015}.
The lens brightness and the extinction values are estimated from the color-color and mass-luminosity relations of main sequence stars \citep{Henry+1993, Kenyon+1995, Kroupa+1997} and the extinction law in \citet{Nishiyama+2009}, respectively. 
We also estimated the source magnitudes in $H$- and $K$-bands from \citet{Kenyon+1995} with taking account of 10\% uncertainty.
Figure \ref{fig:lens_mag} represents the apparent lens magnitudes in each band derived from the Bayesian analysis. 
The dark and light blue regions indicate the 68.3 \% and 95.4\% confidence interval and the vertical blue lines indicate the median values.
The vertical solid and dashed red lines are the source magnitudes and its $1\sigma$ uncertainties in each passband.
The relationship between the heliocentric and geocentric relative proper motion is 
\begin{equation}
\bm{\mu}_{\rm rel, H} = \bm{\mu}_{\rm rel, G} + \frac{\pi_{\rm rel}}{\rm au}\bm{v_{\oplus}}
\end{equation}
where $\pi_{\rm rel}={\rm au}(D^{-1}_L-D^{-1}_S)$ and $\bm{v_{\oplus}}=(v_{\oplus, N}, v_{\oplus, E})=(-2.91, 9.44)\;{\rm km\;s^{-1}} $ are the relative lens-source parallax and the instant velocity of Earth on the plane of the sky at the reference time, respectively.  
The heliocentric relative proper motion is $\mu_{\rm rel, H}\sim2.5\;{\rm mas\;yr^{-1}}$ and thus the angular separation between the source and lens would be $\sim15\;{\rm mas}$ in 2019.
\citet{Bhattacharya+2017} have demonstrated the feasibility of {\it Hubble Space Telescope} follow-up observations to measure the separation between the source and the lens with $12$ mas when the lens is not too much fainter than the source (The current state of technical arts for high angular resolution analysis is detailed in \citet{Bhattacharya+2018}).
Hence, it might benefit from a high-resolution follow-up observation in order to constrain the physical parameters of the lens system.
However, we note that the four degenerate solutions have parallax vectors ${\bm \pi_{\rm E}}$ with amplitudes, directions and uncertainties approximately similar to each other and thus it is unlikely that the degenerate solutions are resolved by high-resolution follow-up observations.

\section{Summary \& Discussion \label{sec:sam_dis}}
We have presented the analysis of the microlensing event OGLE-2013-BLG-0911.
The previous research on the event \citep{Shvartzvald+2016} reported that the lensing anomaly could be explained by a planetary mass ratio, $q\approx3\times10^{-4}$.
From a detailed grid search analysis, however, we found that a binary mass ratio $q\approx3\times10^{-2}$ is preferred over a planetary mass ratio to explain the light curve.
Finally, we conclude that the lens system is an M-dwarf orbited by a massive Jupiter companion at very close ($M_{\rm host}=0.30^{+0.08}_{-0.06}M_{\odot}$, $M_{\rm comp}=10.1^{+2.9}_{-2.2}M_{\rm Jup}$, $a_{\rm exp}=0.40^{+0.05}_{-0.04}{\rm au}$) or wide ($M_{\rm host}=0.28^{+0.10}_{-0.08}M_{\odot}$, $M_{\rm comp}=9.9^{+3.8}_{-3.5}M_{\rm Jup}$, $a_{\rm exp}=18.0^{+3.2}_{-3.2}{\rm au}$) separation.
\vspace{0.1in}

Microlensing light curves generally provide much more precise estimation of the mass ratio rather than that of the absolute lens mass.
\citet{Bond+2004} defined the mass ratio boundary between BDs and planets as $q=0.03$ in order to distinguish planetary and stellar binary (including BD) microlensing events.
For this event, the best-fit mass ratio is slightly above the mass ratio boundary of $q=0.03$.
On the other hand, the median mass of the companion is slightly below the lower limit of BD mass of $13M_{\rm Jup}$. 
Therefore, it is ambiguous to classify the companion as a BD or a planet.
In fact, these boundaries are somewhat arbitrary and it might be nonsense to classify such an ambiguous companion according to the boundaries. 
However, the formation mechanisms for BDs and planets are likely to be different and the object near the boundaries could have been formed by either formation mechanism.
Therefore, it would be very important to probe the distribution of intermediate mass companions of $\sim13M_{\rm Jup}$.
\vspace{0.1in}

Missing the best lens model explanation to the observed microlensing light curve data might have serious impacts on any statistical microlensing analysis incorporating those modeling results.
For instance, \citet{Shvartzvald+2016} suggests that there is a possible BD deficit corresponding to $q\sim10^{-2}$ in their detection-efficiency-corrected mass ratio function.
However, we found OGLE-2013-BLG-0911, which was adopted as a planetary sample in their analysis, would correspond to the position of the BD deficit, which would affect their result to some extent.
The reason why they missed the best solution would be the very small/wide projected separation $s\approx0.2$ or $\approx7$. 
They explored the $s$ parameter space of $0.3 < s < 3$ in their grid search analysis.
It is known that a central caustic size is approximately proportional to not only $q$ but also $s^{2}\;({\rm for}\;s\ll1)$ and $s^{-2}\;({\rm for}\;s\gg1)$ \citep{Chung+2005}.
Therefore, when we model microlensing light curves with perturbations caused by possibly small-size central caustics, we should suspect the possibilities of not only very low-mass but also very close and wide lens companions.
The detection efficiency for companions with such extremely close and wide separation is much lower than that with $s\approx1$ \citep{Suzuki+2016}.  
Hence, even a small number of detections may be important in the statistical analysis. 
\vspace{0.1in}

The successful discovery of the best fit model depends on the initial parameters for the MCMC fitting.
Currently, the initial parameters for modeling binary-lens events are mainly based on the experiences of the modelers or the brute-force with the grid search analysis across the wide range of the parameter spaces.
The systematic analysis of many events relies on the latter method. 
However, it would not work if the best-fit solutions are out of range of the grid search, which happened on this event OGLE-2013-BLG-0911. 
Broadening the search range as possible is a straightforward way to avoid the problem.
However, it is computationally expensive and it is getting more difficult for statistical analysis including hundreds of stellar binary events in the recent high cadence surveys by MOA, OGLE and KMTNet \citep{Kim+2016}.
Furthermore, the {\it Wide Field Infrared Survey Telescope }({\it WFIRST;} \citealp{Spergel+2015}) will be launched in 2025 and be expected to discover $\sim54000$ microlensing events $(|u_0|<3)$ with thousands of binary lens events including $\sim1400$ bound exoplanets with masses of $0.1<M_{p}/M_{\oplus}<10^4$ \citep{Penny+2019}.
We should consider a new method to efficiently search for the best binary-lens solutions.
\citet{Bennett+2012} applied the different parameterization for the wide-separate binary events.
\citet{Khakpash+2019} proposed the algorithm that can rapidly evaluate many binary-lens light curves and estimate the physical parameters of the lens systems, which is successful for very low mass-ratio events but less for higher mass-ratio events.
\vspace{0.1in}

There are only four discoveries of BD companions to M dwarfs within 10 pc from Solar system \citep{Winters+2018}, while approximately 200 M dwarfs are known to exist within 10 pc \citep{Henry+2006,Henry+2016} and much effort has been dedicated to detect such BD companions \citep{Henry+1990,Dieterich+2012}.
Because of their scarcity, incoming new BD discoveries around M dwarfs provide valuable constraints on the formation and evolution theories of stars, BDs and planets. 
Microlensing is a powerful method to probe the BD/massive-planet occurrence frequency across orbital radii $0.1\leq a\leq10$ au around low-mass hosts such as M dwarfs and even BDs \citep{Gaudi2002}, which is challenging for other exoplanet detection methods.
Although microlensing samples generally can not provide some information such as host metallicity and eccentricity, microlensing can provide both thier masses and orbital separations. 
It is very important to uncover the distributions of BD properties by microlensing.

\acknowledgments
We would like to thank the anonymous referee who helped to greatly improve our paper.
The OGLE project has received funding from the National Science Centre, Poland, grant MAESTRO 2014/14/A/ST9/00121 to AU. 
CITEUC is funded by National Funds through FCT - Foundation for Science and Technology (project: UID/Multi/00611/2013) and FEDER - European Regional Development Fund through COMPETE 2020 - Operational Programme Competitiveness and Internationalization (project: POCI-01-0145-FEDER-006922).
D.P.B., A.B., and D.S. were supported by NASA through grant NASA-NNX12AF54G.
Work by C.R. was supported by an appointment to the NASA Postdoctoral Program at the Goddard Space Flight Center, administered by USRA through a contract with NASA. 
Work by N.K. is supported by JSPS KAKENHI grant No. JP15J01676. 
Work by Y.H. is supported by JSPS KAKENHI grant No. JP1702146. 
N.J.R. is a Royal Society of New Zealand Rutherford Discovery Fellow. 
This work was supported by JSPS KAKENHI grant No. JP17H02871. 
This research was supported by the I-CORE program of the Planning and Budgeting Committee and the Israel Science Foundation, grant 1829/12. 
D.M. acknowledges support by the U.S.-Israel Binational Science Foundation.
Work by C.H. was supported by the grant (2017R1A4A1015178) of National Research Foundation of Korea. 
Work by W.Z., Y.K. J., and A.G. was supported by AST1516842 from the US NSF. W.Z., I.G.S., and A.G. were supported by JPL grant 1500811.
Y.T. acknowledges the support of DFG priority program SPP 1992 ``Exploring the Diversity of Extrasolar Planets'' (WA 1047/11-1).
K.H. acknowledges support from STFC grant ST/R000824/1.

\appendix
\section{Calibration for the Source Magnitude\label{ap:calibration}}
We derived the apparent magnitude and color of the source from the measurements of CT13-$I$ and $V$ that were made during the time of high magnification. 
We basically followed the procedure described in \citet{Bond+2017} in order to convert the CT13 instrumental magnitudes into the standard ones.  
We cross-referenced isolated stars around $2^\prime$ of the source between the CT13 catalog reduced by DoPHOT \citep{Schechter+1993} and the OGLE-III catalog \citep{Szymanski+2011}.
We found the following relation as 
\begin{eqnarray}
I_{\rm O3}-I_{\rm CT13} &=& (27.070\pm0.011)-(0.032\pm0.006)(V-I)_{\rm CT13} \nonumber \\
V_{\rm O3}-V_{\rm CT13}&=& (27.851\pm0.017)-(0.101\pm0.011)(V-I)_{\rm CT13}. \nonumber
\end{eqnarray}
Consequently, we obtained the apparent color and magnitude of the source, $(V-I, I)_{S, {\rm CT13}}=(1.904\pm0.009, 19.618\pm0.006)$.
Moreover, we also derived the source color and magnitude from the measurements of OGLE-$I$ and $V$ for confirmation.
We used Equation (1) in \citet{Udalski+2015} to calibrate the OGLE-IV instrumental magnitudes into the standard ones.
We applied $\Delta{\rm ZP}_{I}=-0.056$, $\Delta{\rm ZP}_{V}=0.133$, $\epsilon_{I}=-0.005\pm0.003$ and $\epsilon_{V}=-0.077\pm0.001$ for Equation (1) in \citet{Udalski+2015}, which are obtained by private communication with the OGLE collaboration.
Finally, we derived the apparent source color and magnitude from OGLE-$I$ and $V$, $(V-I,I)_{S,{\rm O4}}=(1.880\pm0.009,19.594\pm0.006)$.

\section{The impact of the assumption for $M_S$ and $D_S$\label{ap:msds}}
\begin{figure*}
\centering
\includegraphics[angle=-90, scale=0.3,clip]{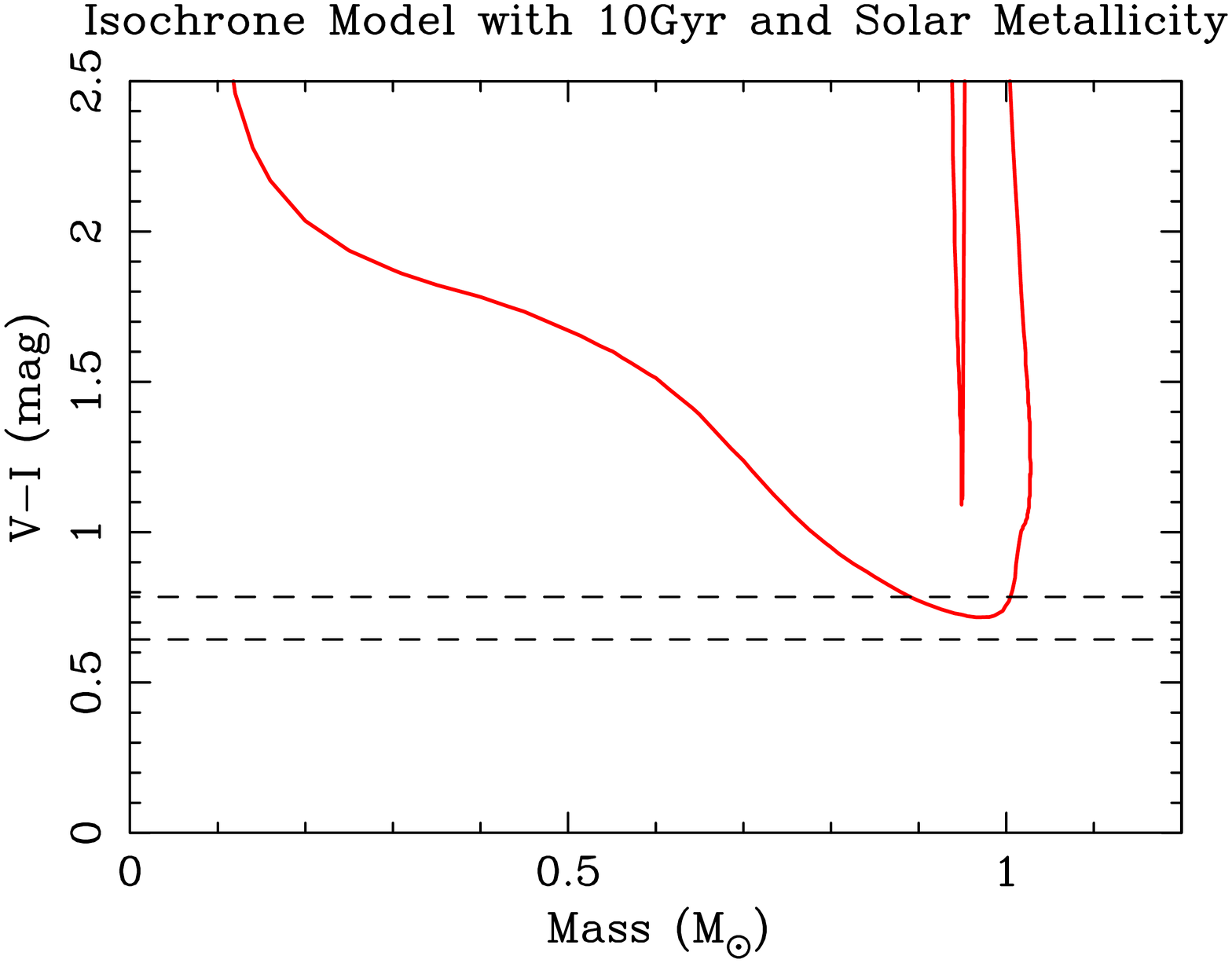}
\includegraphics[angle=-90, scale=0.3,clip]{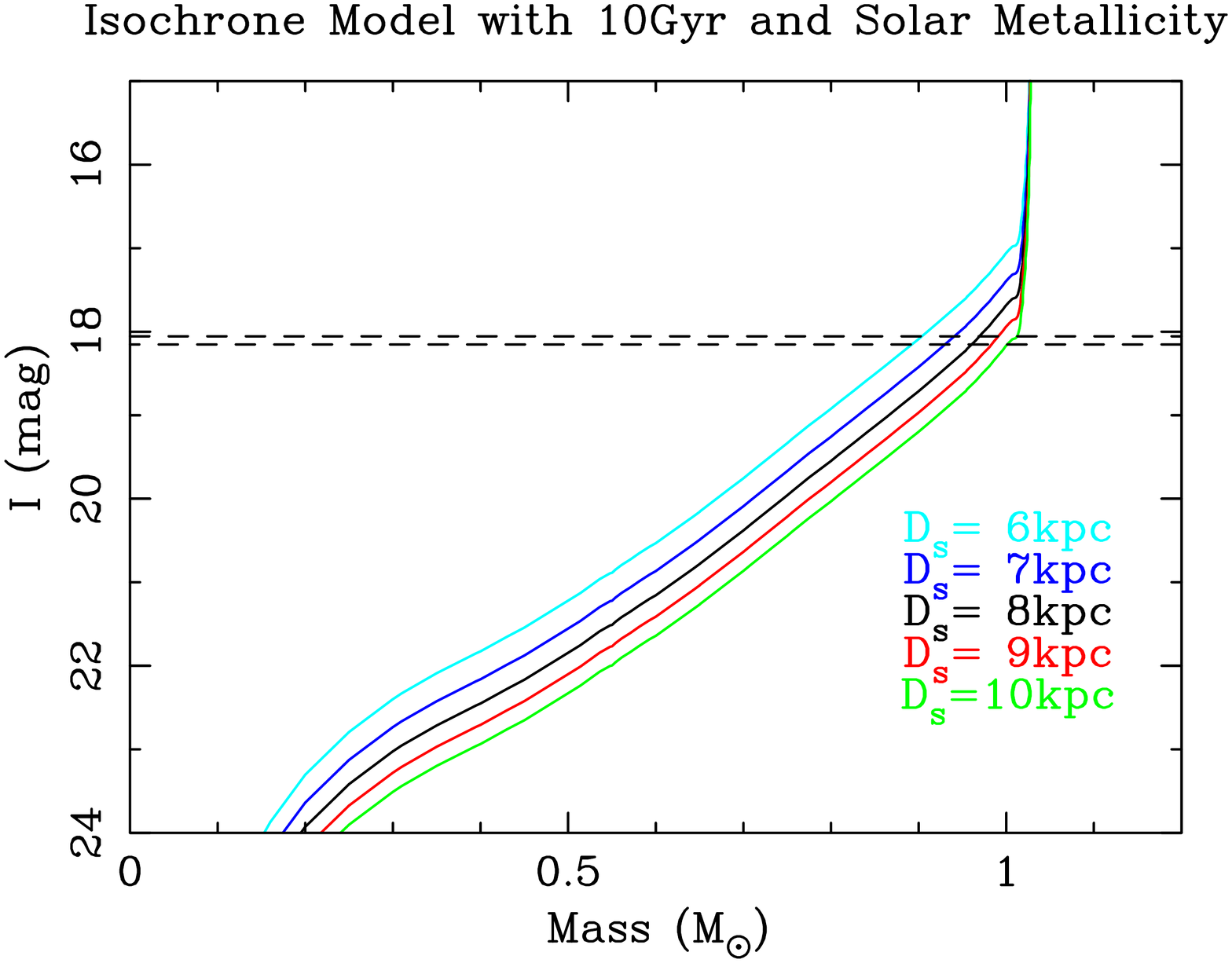}
\caption{
PARSEC stellar isochrone model with solar metallicity and 10 Gyr age.
The areas enclosed by horizontal dashed lines represent the $1\sigma$ ranges for the observed intrinsic source color $(V-I)_{S,0}=0.714\pm0.071$ (right panel) and magnitude $I_{S,0}=18.104\pm0.049$ (left panel), respectively.
}
\label{fig:isochrone}
\end{figure*}
We tested how the assumption of the fixed $M_S=1\;M_{\odot}$ and $D_S=8\;{\rm kpc}$ impact on the final results.
Figure \ref{fig:isochrone} represents the PARSEC stellar isochrone with solar metallicity and 10 Gyr age.
Comparing the isochrone to the observed intrinsic source color and magnitude $(V-I, I)_{S,0}=(0.714\pm0.071,18.104\pm0.049)$, we can state that the source mass and distance are likely to be in the ranges of $0.9\le M_S/M_{\odot}\le 1.0$ and $6\;{\rm kpc}\le D_S \le 10\;{\rm kpc}$, respectively.
In these likely ranges, we conducted light curve modeling for 1L2S, 2L1S and 2L2S with all the 15 combinations of the fixed $M_S=(0.9,0.95,1.0)\;M_{\odot}$ and $D_S=(6,7,8,9,10)$ kpc.
We found that the fixed values have little effects on the best-fit $\chi^2$ value and the MCMC posterior distributions for the lens physical parameters are consistent each other within $1\sigma$ except for the lens distance $D_L$. 
Therefore, we conclude that the assumptions for $M_S$ and $D_S$ do not significantly affect the final results except $D_L$.



\end{document}